\documentclass[preprint,secnumarabic,amssymb, nobibnotes,notitlepage, aps, prd]{revtex4-1}

\usepackage{amsmath, amsthm, amssymb}
\usepackage{tabu}

\usepackage{graphicx}
\usepackage{bm} 

\newcommand{\cov}{\text{cov}}

\newcommand{\R}{{\mathbb R}}

\newcommand{\ee}{\textrm{e}}
\newcommand{\ii}{\textrm{i}}
\newcommand{\xx}{\textrm{x}}
\newcommand{\vvec}[1]{{\bm #1}}

\newcolumntype{+}{!{\vrule width 2pt}}

\newlength\savedwidth

\usepackage{setspace,lipsum}
\usepackage{ragged2e}

\begin{document}
\title{Synaptic Plasticity in Correlated Balanced Networks}%

\author{Alan Eric Akil}%
\affiliation{Department of Mathematics, University of Houston, Houston, Texas, United States of America}

\author{Robert Rosenbaum}%
\affiliation{Department of Applied and Computational Mathematics and Statistics, University of Notre Dame, Notre Dame, Indiana, United States of America}
\altaffiliation{Interdisciplinary Center for Network Science and Applications, University of Notre Dame, Notre Dame, Indiana, United States of America}

\author{ Kre\v{s}imir Josi\'c\ }%
\email[Corresponding author: ]{josic@math.uh.edu}
\affiliation{Department of Mathematics, University of Houston, Houston, Texas, United States of America\\}
\altaffiliation{Department of Biology and Biochemistry, University of Houston, Houston, Texas, United States of America}

\date{April 2020}%
\singlespacing
\begin{abstract}
The dynamics of local cortical networks are irregular, but correlated. Dynamic excitatory--inhibitory balance is a plausible mechanism that generates such irregular activity, but it remains unclear how balance is achieved and maintained in plastic neural networks. In particular, it is not fully understood how plasticity induced changes in the network affect balance, and in turn, how correlated, balanced activity impacts learning. How does the dynamics of balanced networks change under different plasticity rules? How does correlated spiking activity in recurrent networks change the evolution of weights, their eventual magnitude, and structure across the network? To address these questions, we develop a general theory of plasticity in balanced networks. We show that balance can be attained and maintained under plasticity induced weight changes. We find that correlations in the input mildly, but significantly affect the evolution of synaptic weights. Under certain plasticity rules, we find an emergence of correlations between firing rates and synaptic weights. Under these rules, synaptic weights converge to a stable manifold in weight space with their final configuration dependent on the initial state of the network. Lastly, we show that our framework can also describe the dynamics of plastic balanced networks when subsets of neurons receive targeted optogenetic input. 
\end{abstract}

\maketitle

\section{Introduction}
Neuronal activity in the cortex is irregular, correlated, and frequently dominated by a low dimensional component~\cite{Okun2008,Cohen2011,Smith2013,Ecker2014,Tan2014,McGinley2015}. Many early models designed to explain the mechanisms that drive irregular neural activity also resulted in asynchronous states~\cite{Vreeswijk1998,Renart2010}.  However, more recent extensions have shown how correlated states can be generated both internally and exogenously, while preserving irregular single cell activity~\cite{Huang2019,Baker2019,Rosenbaum2017,Landau2018,Mastrogiuseppe2018}. 

Correlated activity can also drive synaptic plasticity~\cite{Morrison2007,Kempter1999}. It is thus important to understand how irregular, correlated cortical activity shapes the synaptic architecture of the network, and in turn, how changes in synaptic weights affect network dynamics and correlations. Neural activity is also characterized by a balance between excitation and inhibition~\cite{Atallah2009,Barral2016,Okun2008,Dehghani2016,Galarreta1998,Yizhar2011,Zhou2014}. We
therefore ask how weights and population dynamics evolve in such states, and whether different types of synaptic plasticity drive a network out of balance, or help maintain balance in the presence of correlations? 

To address these questions, we develop a general theory that relates firing rates, spike count covariances, and changes in synaptic weights in  balanced spiking networks.  Our framework allows us to analyze the effect of general pairwise spike--timing dependent plasticity (STDP) rules. We show how the weights and the network's dynamics evolve under different classical rules, such as Hebbian plasticity, Kohonen's rule, and a form of inhibitory plasticity~\cite{Hebb1949,Markram1997,Bi1998,Kohonen1984,Vogels2011}. 
In general, the predictions of our theory agree well with empirical simulations.  We also explain when and how mean field theory disagrees with simulations, and develop a semi--analytic extension of the theory that explains these disagreements.  Our theory thus allows us to understand mean synaptic weight dynamics, and predicts how synaptic plasticity impacts network dynamics. 

We find that spike train correlations, in general, have a mild effect on the synaptic weights and firing rates. Correlations can affect synaptic weights in two different ways: shifting the stable fixed point of synaptic weights or modulating the speed at which this equilibrium is reached. 
We show that for some rules, synaptic competition can introduce correlations between synaptic weights and firing rates. Such correlations are commonly ignored in the balanced network literature and  can lead to the formation of a stable manifold of fixed points in weight space. This yields a system that realizes different equilibrium synaptic weights depending on the initial weight distribution. 
Lastly, we apply our theory to show how inhibitory STDP~\cite{Vogels2011} can lead to a reestablishment of a balanced state that is broken by optogenetic stimulation of a neuronal subpopulation~\cite{Ebsch2018}. We thus extend the classical theory of balanced networks to understand how synaptic plasticity shapes their dynamics.

\section{Materials and Methods}
\subsection{Review of mean--field theory of balanced networks}
In mammals, local cortical networks can be comprised of thousands of cells, with each neuron receiving thousands of inputs from cells within the local network, and other cortical layers, areas, and thalamus~\cite{Binzegger2004}. Predominantly excitatory, long--range inputs would lead to high, regular firing unless counteracted by local inhibition. To reproduce the sparse, irregular activity observed in cortex,  model networks often exhibit a balance between excitatory and inhibitory inputs \cite{Vreeswijk1996,Vreeswijk1998,Renart2010,Baker2019,Ahmadian2013,Hennequin2018,Doiron2014, Rosenbaum2014}.
This balance can be achieved robustly and without tuning, when synaptic weights are scaled like $\mathcal O(1/\sqrt{N})$, where $N$ is the network size \cite{Vreeswijk1996,Vreeswijk1998}. In this balanced state mean excitatory and inhibitory inputs cancel one another, and the activity is asynchronous~\cite{Renart2010}. Inhibitory inputs can also track excitation at the level of cell pairs, cancelling each other in time, and produce a correlated state~\cite{Baker2019,Okun2008}. 

We first review the mean--field description of  asynchronous and correlated states in balanced networks, and provide expressions for firing rates and spike count covariances averaged over subpopulations that accurately describe  networks of more than a few thousand neurons~\cite{Vreeswijk1996,Vreeswijk1998,Rosenbaum2014,Landau2016,Pyle2016, Baker2019,Rosenbaum2017}: Let $N$ be the total number of neurons in a recurrent network composed of $N_\ee$ excitatory and $N_\ii$ inhibitory neurons. Cells in this recurrent 
network also receive input from $N_\xx$ external Poisson neurons firing at rate $r_\xx$, and with pairwise correlation $c_\xx$ (See Fig. \ref{fig:Kohonen}~\textbf{\textsf{A}}, and Appendix~\ref{rev_meanfield} for more details).
We assume that $q_b=N_b/N \sim \mathcal O(1)$ for $b=\ee,\ii,\xx$. Let $p_{ab}$ be the probability of a synaptic connection,  and  $j_{ab}\sim \mathcal O(1)$ the weight of a synaptic connection from a neuron in population $b=\ee,\ii,\xx$ to a neuron in population $a=\ee,\ii,\xx$.  For simplicity we assume that both the probabilities, $p_{ab},$ and weights, $j_{ab}\sim \mathcal O(1)$ are constant across pairs of subpopulations, although this assumption is not essential.  We define the recurrent, and feedforward mean--field connectivity matrices as

$$
\overline W=\left[\begin{array}{cc}\overline w_{\ee\ee} & \overline w_{\ee\ii}\\ \overline w_{\ii \ee} & \overline w_{\ii\ii}\end{array}\right], \qquad \textrm{ and  } \qquad  \overline W_x=\left[\begin{array}{c}\overline w_{\ee\xx} \\ \overline w_{\ii \xx} \end{array}\right],
$$

\noindent where $\overline w_{ab}=p_{ab}j_{ab}q_b\sim\mathcal O(1)$. 

Let $\vvec r=[r_\ee ,\ r_\ii]^T$ be the vector of mean excitatory and inhibitory firing rates. 
The mean external input and recurrent input to a cell are then
$
\overline{\vvec X}=\sqrt N \overline W_\xx r_\xx,
$
and 
$
\overline{\vvec R}=\sqrt N \overline W \vvec r,
$ 
respectively, and the mean total synaptic input to any neuron is given by

$$
\overline{\vvec I}=\sqrt N\left[\overline W \vvec r+\overline W_\xx r_\xx\right].
$$

We next make the ansatz that in the balanced state the mean input and firing rates remain finite as the network grows, \emph{i.e.}  $\overline{\vvec I}, \vvec r \sim \mathcal O(1)$~\cite{Vreeswijk1996, Vreeswijk1998, Rosenbaum2014,Landau2016,Pyle2016, Baker2019}. This is only achieved when external and recurrent synaptic inputs are in balance, that is when
 
 \begin{equation}\label{eq:rbal}
\lim_{N\to\infty} \vvec r=-\overline W^{-1}\overline W_\xx r_\xx
\end{equation}

\noindent provided that also $\overline X_e /\overline X_i>w_{ei}/w_{ii}>w_{ee}/w_{ie}$ \cite{Vreeswijk1996,Vreeswijk1998}. Eq.~(\ref{eq:rbal}) holds in both the asynchronous and correlated states. 

We define the mean spike count covariance matrix as:  

$$
C=\left[\begin{array}{cc}C_{\ee\ee} & C_{\ee\ii}\\ C_{\ii \ee} & C_{\ii\ii}\end{array}\right]
$$

\noindent where $C_{ab}$ is the mean spike count covariance between neurons in populations $a=\ee,\ii$ and $b=\ee,\ii$ respectively.
From~\cite{Rosenbaum2017, Baker2019}  it follows that in large networks, to leading order in $1/N$ (See \cite{Tetzlaff2012,Grytskyy2013,Helias2014,Renart2010} for similar expressions derived for similar models), 

\begin{align} \label{eq:eqCorr}
C\approx \frac{1}{N}T_{\rm win}\overline W^{-1} \Gamma \overline W^{-T} 
- \frac{1}{N} 
\begin{bmatrix} 
\frac{r_{\ee}T_{\rm win} F_\ee}{q_\ee} & 0 \\
0 & \frac{r_{\ii}T_{\rm win} F_\ii}{q_\ii}
\end{bmatrix}.
\end{align} 

\noindent In Eq.~(\ref{eq:eqCorr}), $F_a$ is the Fano factor of the spike counts averaged over neurons in populations $a=\ee,\ii$ over time windows of size $T_{\rm win}$. The second term in Eq.~(\ref{eq:eqCorr}) is $\mathcal O (1/N)$ and accounts for intrinsically generated variability~\cite{Baker2019}.
The matrix $\Gamma$ has the same structure as $C$ and represents the covariance between external inputs. 
If external neural activity is uncorrelated ($c_\xx=0$), then 

$$
\Gamma=\overline W_\xx \overline W_\xx^T\frac{r_\xx}{q_\xx}\sim\mathcal O(1)
$$

\noindent so that $C\sim\mathcal O(1/N)$, and the network is in an \emph{asynchronous} regime. 
If external neural activity is correlated with mean pairwise correlation coefficient $c_\xx \neq 0$, then in leading order $N$, 

$$
\Gamma=N\overline W_\xx \overline W_\xx^T c_\xx r_\xx\sim\mathcal O(N),
$$

\noindent so that $C\sim\mathcal O(1)$, and the network is in a \emph{correlated} state. 
Eq.~(\ref{eq:eqCorr}) can be extended to cross--spectral densities as shown in Appendix~\ref{rev_meanfield}\ref{cov_meanfield} and by \emph{Baker et al.}~\cite{Baker2019}.

\subsection{Network model}
For illustration, we used recurrent networks of $N$ exponential integrate--and--fire (EIF) neurons (See  Appendix~\ref{rev_meanfield}), $80\%$ of which were excitatory ($E$) and $20\%$ inhibitory ($I$)~\cite{Baker2019,LitwinKumar2014,Ebsch2018,Rosenbaum2017,FourcaudTrocme2003}.
The initial connectivity structure was random:

$$
J_{jk}^{ab}=\frac{1}{\sqrt N}\begin{cases}j_{ab} & \textrm{with probability } p_{ab}, \\ 0 & \textrm{otherwise}.\end{cases}
$$

\noindent Initial synaptic weights were therefore independent.  We set $p_{ab} = 0.1$ for all $a,b=\ee,\ii$, and denote by $J^{ab}_{jk}$ the  weight of a synapse between presynaptic neuron $k$ in population $b=\ee,\ii,\xx$ and  postsynaptic neuron $j$ in population $a=\ee,\ii,\xx$. We modeled postsynaptic currents using an exponential kernel,  $K_a(t)=\tau_{a}^{-1}e^{-t/\tau_{a}}H(t)$ for each $a=\ee,\ii,\xx$ where $H(t)$ is the Heaviside function.

\noindent
\subsubsection{ Synaptic plasticity rules } 

To  model activity--dependent changes in synaptic weights we used eligibility traces to define the propensity of a synapse to change~\cite{Klopf1982,Houk1995,Izhikevich2007,He2015,gerstner2018}. The eligibility trace, $x_j^a(t)$, of neuron $j$ in population $a$ evolves according to

\begin{equation} \label{eq:trace}
\tau_{\mathrm{STDP}}\frac{dx_j^a (t) }{dt}=-x_j^a (t) +\tau_{\mathrm{STDP}} S_j^a(t),
\end{equation}

\noindent for $a=\ee,\ii$, where  $S_j^a(t)=\sum_n \delta(t-t_n^{a,j})$ is the sequence of spikes of  neuron, $j$. The eligibility trace, and the time constant, $\tau_{\mathrm{STDP}}$ define a period following a spike in the pre-- or post--synaptic cell during which a synapse can be modified by a spike in its counterpart.

Our general theory of synaptic plasticity allows any synaptic weight to be subject to constant drift, changes due to pre-- or post--synaptic activity only, and/or due to pairwise interactions in activity between the pre-- and post--synaptic cells (zero, first, and second order terms, respectively, in Eq.~(\ref{eq:dJ})). The theory can be extended to account for other types of interactions.
Each synaptic weight therefore evolves according to a generalized STDP rule:

\begin{align} \label{eq:dJ}
\frac{dJ^{ab}_{jk}}{dt}= \eta_{ab} \bigg(
A_0 +
\sum_{ \alpha=\{a,j\},\{b,k\}} A_{\alpha} S_\alpha +
\sum_{ \alpha,\beta=\{a,j\},\{b,k\}} B_{\alpha,\beta} x_\alpha S_\beta  \bigg)
\end{align}

%

\noindent where $\eta_{ab}$ is the learning rate that defines the timescale of synaptic weight changes, $A_0,A_\alpha,B_{\alpha\beta}$ are functions of the synaptic weight, $J^{ab}_{jk},$ and $a,b=\ee,\ii$. 
For instance, the term $B_{(\ee,k),(\ii,j)}  x_k^\ee S_j^\ii$ represents the contribution due to a spike in post--synaptic cell $j$ in the inhibitory subpopulation, at the value  $x_k^\ee$ of the eligibility trace in the pre--synaptic cell $k$ in the excitatory subpopulation. Only pairwise interactions between spikes and eligibility traces are meaningful, as the probability that two spikes occur at the same time is negligible.
We used a single timescale for all synapses, but the theory can be generalized.

  \begin{table}[h!]
  \begin{center}
 \begin{tabular}{|p{0.3\textwidth}|p{0.3\textwidth}|p{0.4\textwidth}|} 
 \hline
 STDP Rule & Coefficients & Equation 
 \\
 \hline
 Classical $EE$\newline Hebbian~\cite{Hebb1949,Markram1997,Bi1998}
 & $B_{(\ee,j),(\ee,k)} = -J_{jk}^{\ee\ee}$ \newline $B_{(\ee,k),(\ee,j)} = J_{\rm max}$ 
 & \ \newline $\frac{dJ_{jk}^{\ee\ee}}{dt} = \eta_{\ee\ee} \big( J_{\rm max}x_k^\ee S_j^\ee - J_{jk}^{\ee\ee}x_j^\ee S_k^\ee \big)$ \newline 
 \\
 \hline
 Classical $EE$ \newline Anti--Hebbian~\cite{Markram1997,Bi1998}
 & $B_{(\ee,j),(\ee,k)} = J_{jk}^{\ee\ee}$ \newline $B_{(\ee,k),(\ee,j)} = - J_{\rm max}$ 
 & \ \newline $\frac{dJ_{jk}^{\ee\ee}}{dt} = \eta_{\ee\ee} \big( - J_{\rm max}x_k^\ee S_j^\ee + J_{jk}^{\ee\ee}x_j^\ee S_k^\ee \big)$ \newline \ 
 \\
 \hline
 Homeostatic Inhibitory ~\cite{Vogels2011}
 & $A_{\ii,k}= \alpha_{\ee} J_{jk}^{\ee\ii}/J^{\ee\ii}_0 $
 \newline $B_{(\ee,j),(\ii,k)}=- J_{jk}^{\ee\ii}/J^{\ee\ii}_0$ 
 \newline $B_{(\ii,k),(\ee,j)}=- J_{jk}^{\ee\ii}/J^{\ee\ii}_0$ 
 &\ \newline $
\frac{dJ_{jk}^{\ee\ii}}{dt}= - \eta_{\ee\ii} \frac{J_{jk}^{\ee\ii}}{J^{\ee\ii}_0} \left[ (x_j^\ee-\alpha_{\ee})S_k^\ii+ x_k^\ii S_j^\ee \right]
$ \newline \
 \\
 \hline
 Oja's Rule~\cite{Oja1982}
 & $B_{(\ee,j),(\ee,j)}=-J_{jk}^{\ee\ee} $
 \newline $B_{(\ee,j),(\ee,k)}=\beta$ 
 & \ \newline $\frac{dJ_{jk}^{\ee\ee}}{dt}= \eta_{\ee\ee} \big( \beta x_j^\ee S_k^\ee - J_{jk}^{\ee\ee} x_j^\ee S_j^\ee \big)$ \newline \ 
 \\
 \hline
 Kohonen's Rule~\cite{Kohonen1984}
 & 
 $A_{\ee,j}=-J_{jk}^{\ee\ee} $
 \newline $B_{(\ee,j),(\ee,k)}=\beta $ 
 \ 
 & \ \newline \ 
 $\frac{dJ_{jk}^{\ee\ee}}{dt}= \eta_{\ee\ee} \big( \beta x_j^\ee S_k^\ee - J_{jk}^{\ee\ee} S_j^\ee \big)$ \newline \  
 \\
 \hline
\end{tabular}
\end{center}
\caption{\textbf{Examples of STDP rules.} A number of different plasticity rules can be obtained as special cases of the general form given in Eq.~(\ref{eq:dJ}).
} \label{table:STDPrules}
\end{table}

This general formulation captures a range of classical plasticity rules as special examples: Table \ref{table:STDPrules} shows that different choices of parameters yield Hebbian~\cite{Hebb1949,Markram1997,Bi1998}, anti--Hebbian, as well as Oja's~\cite{Oja1982}, and other rules.
We only considered interactions up to second order. The BCM rule~\cite{BCM1982}, and other rules~\cite{Pfister2006,Gjorgjieva2011} that depend on interactions beyond second order will be considered elsewhere.

\subsection{Dynamics of mean synaptic weights in balanced networks}

To understand how the dynamics of the network, and synaptic weights co--evolve we derived effective equations for the firing rates, spike count covariances, and synaptic weights using Eqs.~(\ref{eq:rbal}--\ref{eq:eqCorr}). The following is an outline, and details can be found in Appendix~\ref{effect_corrs}. 

We assumed that changes in synaptic weights occur on longer timescales than the dynamics of the eligibility traces and the correlation timescale, \emph{i.e.} $1/\eta_{ab} \gg T_{\rm win}$ ~\cite{Gilson2009I,Gilson2009II,Gilson2009III,Gilson2009IV,Gilson2010V,Kempter1999}. 
Let  $\Delta T$  be a time increment larger than the timescale of eligibility traces, $\tau_{\mathrm{STDP}},$ and $T_{\rm win}$, but smaller than $1/\eta_{ab}$, so that the difference quotient of the weights and time is given by~\cite{Kempter1999}:

\begin{align}
\label{E:averaging}
\frac{\Delta J^{ab}_{jk}}{\Delta T} & = \frac{\eta_{ab}}{\Delta T} \int_0^{\Delta T} \Big[  A_0 + \sum_{ \alpha=\{a,j\},\{b,k\}} A_{\alpha} S_\alpha +
\sum_{ \alpha,\beta=\{a,j\},\{b,k\}} B_{\alpha,\beta} x_\alpha S_\beta    \Big] dt.
\end{align}

\noindent The difference in timescales allows us to assume that the firing rates and covariances are in quasi--equilibrium. 
Replacing the terms on the right hand side of Eq.~\eqref{E:averaging}, with
their averages over time, and over different network subpopulations, we obtain the following mean--field equation for the weights:

\begin{align} \label{eq:dJsoln} 
\frac{d J_{ab}}{dt}  = \eta_{ab} \bigg( A_0 +  \sum_{\alpha,\beta=\{a,b\}}  \textrm{Rate}_{\alpha,\beta} + \textrm{Cov}_{\alpha,\beta} \bigg) ,
\end{align}

\noindent where

\begin{align*} 
\textrm{Rate}_{\alpha,\beta} &= A_{\alpha} r_{\alpha}/2 + B_{\alpha,\beta} \tau_{STDP}r_\alpha r_\beta, \\
\textrm{Cov}_{\alpha,\beta} &= B_{\alpha,\beta} \int_{-\infty}^\infty \widetilde K(f)\langle S_\alpha, S_\beta\rangle(f)df,
\end{align*}

\noindent and $\widetilde K(f)$ is the Fourier transform of the synaptic kernel, $K(t)$. 
Recall that $\langle S_\alpha,S_\beta \rangle (f)$ is the average cross spectral density of spike trains in populations $\alpha,\beta$. The cross spectral density (CSD) of a pair of spike trains is defined as the Fourier transform of the covariance function between the two spike trains, and when evaluated at $f=0$, the CSD is proportional to the spike count covariance between the two spike trains (See Appendix~\ref{rev_meanfield}\ref{cov_meanfield}). 

For example, classical Hebbian \emph{EE} plasticity in Table \ref{table:STDPrules} leads to the following mean--field equation,

$$
\frac{dJ_{\ee\ee}}{dt}= \eta_{\ee\ee} \bigg(J_{\rm max}-J_{\ee\ee} \bigg) \bigg(\tau_{\rm STDP}r_{\ee}^2 +\int_{-\infty}^\infty \widetilde K(f) \langle S_{\ee},S_{\ee} \rangle (f) df \bigg). 
$$

\noindent Eqs.~(\ref{eq:rbal},\ref{eq:eqCorr},\ref{eq:dJsoln}) thus self--consistently describe the macroscopic dynamics of the balanced network. These three equations are coupled, and can be solved by first finding the firing rates and covariances (both depend on the plastic weight $J_{ab}$) obtained using the mean--field description of balanced networks, Eqs.~(\ref{eq:rbal}--\ref{eq:eqCorr}), substituting the results into Eq.~(\ref{eq:dJsoln}), and then finding the roots.
We can then use the synaptic weights (root of Eq.~(\ref{eq:dJsoln})) to obtain the corresponding rates and covariances using Eqs.~(\ref{eq:rbal}--\ref{eq:eqCorr}). This process can be repeated iteratively at each time step to obtain the evolution and equilibrium of firing rates, spike count covariances, and synaptic weights provided our assumption about the separation of timescales holds.

\subsection{Perturbative analysis}

We next show how rates and spike count covariances are impacted by perturbations in synaptic weights.
At steady state the average firing rates in a balanced network with mean--field connectivity matrix $\overline W_0$ 
are given by

$$
\vvec r_0 = -\overline W_0^{-1} \overline W_\xx r_\xx
$$

\noindent We assume that the mean--field connectivity matrix is perturbed to $\overline W_{\rm perturb} = \overline W_0 + \Delta \overline W
$.  Using Neumann's approximation~\cite{Kincaid2002},
$(I+ H)^{-1} \approx (I- H)$, which holds for any square matrix $H$ with $\left\lVert H\right\rVert < 1$, and 
ignoring terms of power 2 and larger, we obtain,

\begin{align*}
\overline W_{\rm perturb}^{-1} &= (\overline W_0+\Delta \overline W)^{-1} 
= \big( \overline W_0 (I+ \overline W_0^{-1}\Delta\overline  W)\big)^{-1} \\
& \approx \big( I - \overline W_0^{-1}\Delta \overline W \big) \overline W_0^{-1},
\end{align*}

\noindent where $I$ is the identity matrix of appropriate size.
We use this approximation of the perturbed weights to estimate the rates and spike count covariances using Eqs.~(\ref{eq:rbal}--\ref{eq:eqCorr}).

\subsection{Comparison of theory with numerical experiments}

We define spike trains of individual neurons in the population as sums of Dirac delta functions, $ S_i(t) =\sum_j\delta(t-t_{ij})$,  where the $t_{ij}$ is the time of the $j^{th}$ spike of neuron $i$. Assuming the system has reached equilibrium, we partition the 
interval over which activity has been measured into $K$ equal subintervals, and define the spike count covariance between two cells as,

$$
{\rm{cov}}(n_{1k},n_{2k})= \sum_k (n_1^k - \overline n_1)(n_2^k - \overline n_2),
$$

\noindent where $n_{ik}$ is the spike count of neuron $i$ in subinterval, or time window, $k$, and $\overline n_i = \frac{1}{K} \sum_k n_{ik}$ is the average spike count over all subintervals. In simulations we used subintervals of size $T_{\rm win}=250$ms, although the theory applies to sufficiently long subintervals, and can be extended to shorter intervals as well.  The spike count covariance thus captures shared fluctuations in firing rates between the two neurons~\cite{Doiron2016}.


\section{Results}

To show how  weights and activity co--evolve in correlated balanced networks, we describe the particular case of Kohonen's STDP rule~\cite{Kohonen1984}, acting on excitatory synaptic weights. Our theory predicts that a stable balanced state is realized, and provides asymptotic values for the rates, spike count covariances, and synaptic weights. Simulations agree with these predictions. We next show that correlated activity can have a moderate, but significant impact on the dynamics of synaptic weights. Our unadjusted theory does not correctly predict weight dynamics  for certain plasticity rules when heterogeneity in rates can lead to competition between synaptic weights. The resulting correlations between rates and weights are inconsistent with classical balanced network theory. We show how our framework can be extended to describe such networks as well. Finally, we show that inhibitory STDP is robust to perturbations: it restores balance in a network receiving targeted input to a subset of cells, and returns to previous network states after stimulus offset.

\subsection{Balanced networks with excitatory plasticity}

Excitatory plasticity plays a central role in theories of learning, but can lead to instabilities~\cite{Markram1997,Bi1998,Ocker2015,Babadi2013}.   Our  theory explains the effect of such plasticity rules in balanced networks, and predicts the stability of the balanced state, the fixed point of the system, and the impact of the plasticity rule on the dynamics of the network.

We consider a network in a correlated state with excitatory--to--excitatory (\emph{EE}) weights that evolve according to Kohonen's rule~\cite{Kohonen1984}.  This rule was first introduced in artificial neural networks~\cite{Kohonen1990}, and was later shown to lead to the formation of self--organizing maps in model biological networks. 
We use our theory to show that Kohonen's rule leads to stable asynchronous or correlated balanced states, and verify these predictions in simulations. 


\begin{figure}[!htb]
\centering
  \includegraphics[width=\textwidth]{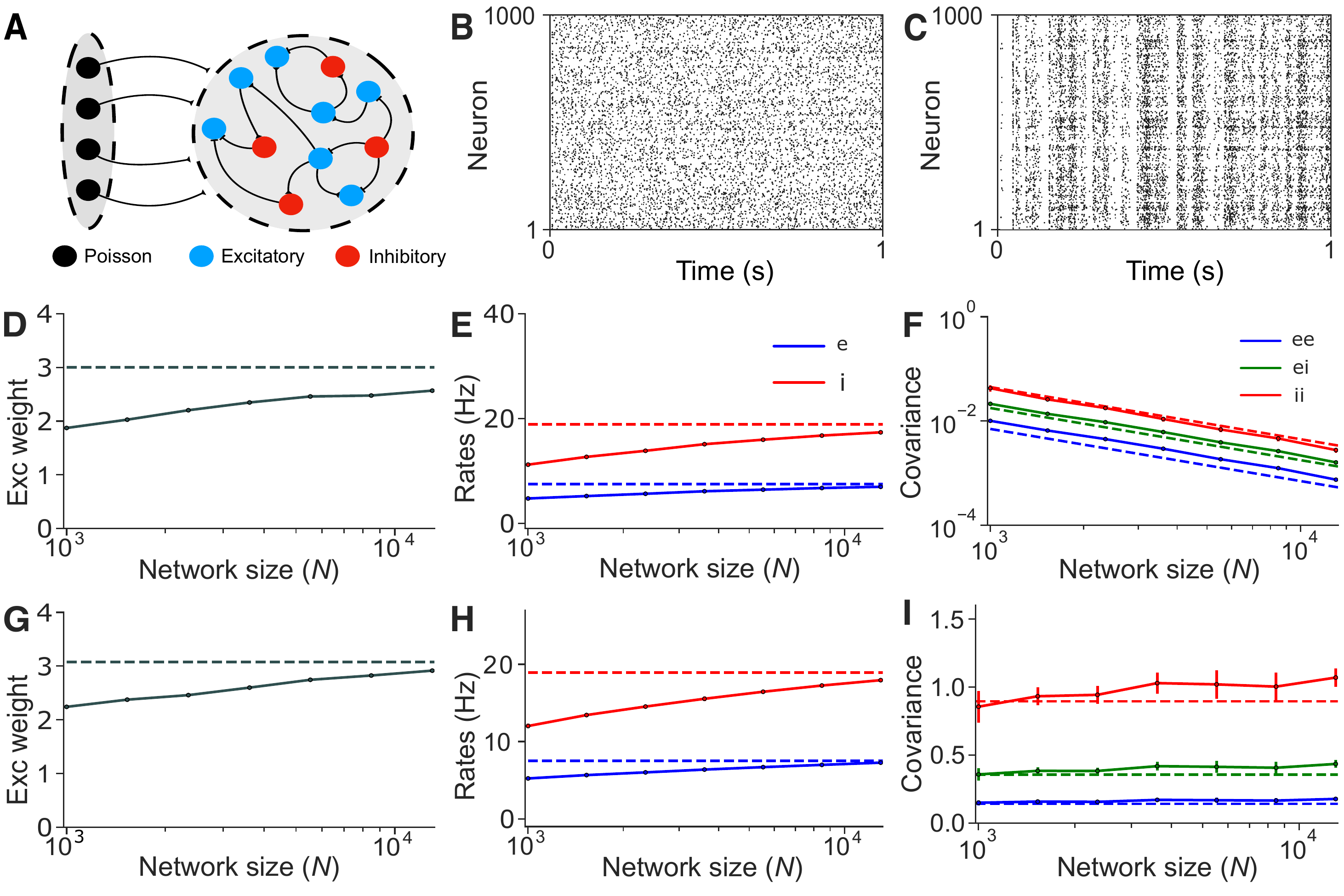}
\caption{ \textbf{A plastic, balanced network in asynchronous and correlated regimes.} \textbf{\textsf{A:}} A recurrent network of excitatory, $E$, and inhibitory, $I$, neurons is driven by an external feedforward layer, $X,$ of correlated Poisson neurons. 
\textbf{\textsf{B:}} Raster plot of 1000 neurons in a network of $N = 10^4$ neurons in an asynchronous state. 
\textbf{\textsf{C:}} Same as (\textbf{\textsf{B}}), but in a correlated state.
\textbf{\textsf{D:}} Mean steady state \emph{EE}  synaptic weight, $j_{ee}$, in an asynchronous state.
\textbf{\textsf{E:}} Mean $E$ and $I$ firing rates for different  network sizes, $N$, in an asynchronous state.
\textbf{\textsf{F:}} Mean \emph{EE, II} and \emph{EI} spike count covariances in an asynchronous state. 
\textbf{\textsf{G--I:}} Same as (\textbf{\textsf{D--F}}) but for a network in a correlated state. 
Solid lines represent simulations, and dashed lines are values obtained using Eqs.~(\ref{eq:rbal},\ref{eq:eqCorr},\ref{eq:dJsoln}). All empirical results were averaged over 10 realizations. In the asynchronous state $c_{\rm x}=0$, and in the correlated state $c_{\rm x}=0.1$. }
\label{fig:Kohonen}
\end{figure}

Kohonen's Rule can be implemented by letting  \emph{EE} synaptic weights evolve according to~\cite{Kohonen1984} (See Table \ref{table:STDPrules}), 

\begin{align} \label{eq:Jee}
\frac{dJ_{jk}^{\ee\ee}}{dt}= \eta_{\ee\ee} \big( \beta x_j^\ee S_k^\ee - J_{jk}^{\ee\ee} S_j^\ee \big) ,
\end{align}

\noindent where $\beta>0$ is a parameter that can change the fixed point of the system. This STDP rule is competitive as weight updates only occur when the pre--synaptic neuron is active, so that the most active pre--synaptic neurons change their synapses more than less active pre--synaptic cells.

The mean--field approximation describing the evolution of synaptic weights given in Eq.~\eqref{eq:dJsoln} has the form:

\begin{align} \label{eq:JeeSoln}
\frac{ dJ_{\ee\ee}}{dt}= & \eta_{\ee\ee} \bigg( \beta \tau_{STDP} r_\ee^2 - J_{\ee\ee} r_\ee + \beta \int_{-\infty}^\infty \widetilde K(f) \langle S_\ee, S_\ee\rangle df \bigg).
\end{align}

\noindent The fixed point of Eq.~(\ref{eq:JeeSoln}) can be obtained by using the expressions for the rates and covariances obtained in the balanced state (Eqs.~(\ref{eq:rbal}--\ref{eq:eqCorr})). The rates and covariances at steady--state can then be obtained from the resulting weights.  
 
Our theory predicts that the network attains a stable balanced state, and the average rates, weights, and covariances at this equilibrium (Fig. \ref{fig:Kohonen}). 
These predictions agree with numerical simulations in both the asynchronous and correlated states (Fig. \ref{fig:Kohonen} \textbf{\textsf{B}},\textbf{\textsf{C}}). 
As expected, predictions improve with network size, $N$, and spike count covariances scale as $1/N$ in the asynchronous state (Fig. \ref{fig:Kohonen} \textbf{\textsf{D}}--\textbf{\textsf{F}}). 
Similar agreement holds in the correlated state, including the impact of the correction introduced in Eq.~(\ref{eq:JeeSoln}) (Fig. \ref{fig:Kohonen} \textbf{\textsf{G}}--\textbf{\textsf{I}}).

The predictions of the theory hold in all cases we tested (See Appendix~\ref{more_res_disc}\ref{hebbian_results} for a network under Classical Hebbian STDP). Understanding when 
plasticity will support a stable balanced state allows one to implement Kohonen's rule in complex contexts and tasks, without the
emergence of pathological states (\emph{e.g.} unconstrained Classical Hebbian STDP can lead to divergence of excitatory firing rates due to runaway potentiation of \emph{EE} synapses).


\begin{figure}[!htb]
\centering
  \includegraphics[width=\textwidth]{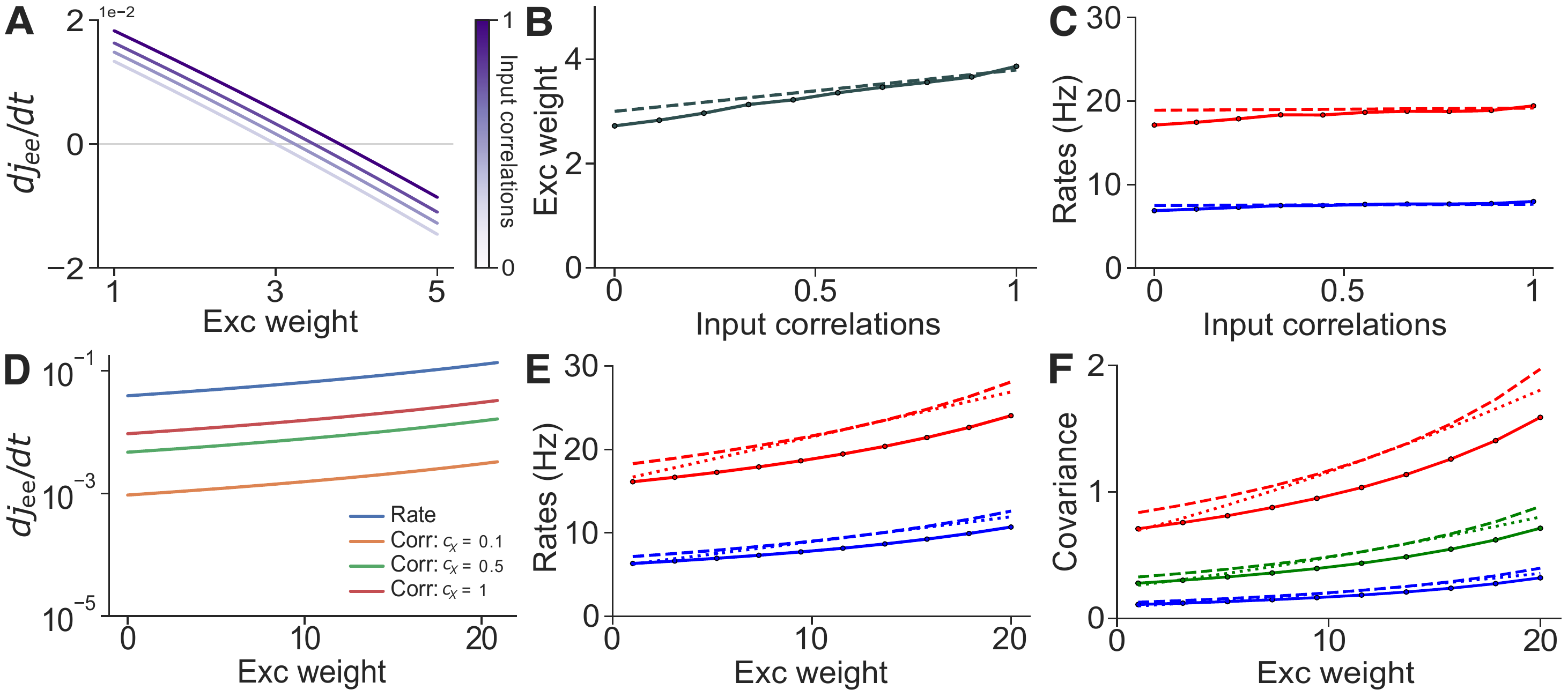}
\caption{\textbf{Spike count covariances mildly impact the fixed point of synaptic weights and firing rates.} \textbf{\textsf{A:}} The rate of change of \emph{EE} weights as function of the weight,  $j_{\ee\ee},$ at different levels of input correlations, $c_{\xx}$.
\textbf{\textsf{B:}} Mean steady--state \emph{EE} synaptic weight for a range of input correlations, $c_{\xx}$.
\textbf{\textsf{C:}}  Mean $E$ and $I$ firing rates as a function of input correlations.
\textbf{\textsf{D:}} Same as (\textbf{\textsf{A}}) but for an \emph{EE} STDP rule with all coefficients involving order 2 interactions set equal to 1, and all other coefficients set equal to zero.
 \textbf{\textsf{E:}} Mean $E$ and $I$ firing rates as a function of mean $EE$ synaptic weights. \textbf{\textsf{F:}} Mean spike count covariances between $E$ spike trains, $I$ spike trains, and between $E$--$I$ spike trains as a function of $EE$ synaptic weight, $j_{\ee\ee}$. Solid lines represent simulations (except in \textbf{\textsf{A}},\textbf{\textsf{D}}), dashed lines are values obtained from theory (Eqs.~(\ref{eq:rbal},\ref{eq:eqCorr},\ref{eq:dJsoln})), and dotted lines were obtained from the perturbative analysis. Note that in all panels, `Exc weight' refers to $j_{\ee\ee}$ rather than $J_{\ee\ee}$, as the former does not depend on $N$.}
\label{fig:DoCorrsMatterKohonen}
\end{figure}

\subsubsection{Dynamics of correlated balanced networks under excitatory STDP}
We next asked whether and how the equilibrium and its stability are affected by correlated inputs to a plastic balanced network. In particular, we used our theory to determine whether changes in synaptic weights are driven predominantly by the firing rates of the pre-- and post--synaptic cells, or correlations in their activity. We also asked whether correlations in neural activity can change the equilibrium, the speed of convergence to the equilibrium, or both? 


We first address the role of correlations. As shown in the previous section. Our theory predicts that  a plastic balanced network remains stable under Kohonen's rule, and that an increase in the mean \emph{EE} weights by $10-20\%$ as correlations in the input are increased. Both predictions were confirmed by simulations (Fig. \ref{fig:DoCorrsMatterKohonen} \textbf{\textsf{A}},\textbf{\textsf{B}}). The theory also predicted that this increase in synaptic weights results in negligible changes in firing rates, which simulations again confirmed (Fig. \ref{fig:DoCorrsMatterKohonen} \textbf{\textsf{C}}). 

How large is the impact of correlations in plastic balanced networks more generally? To address this question, we assumed that only pairwise interactions affect \emph{EE} synapses, as first order interactions depend only on rates after averaging. We thus set $B_{\alpha,\beta}\equiv 1$, and all other coefficients to zero in Eq.~(\ref{eq:dJ}). While the network does not reach a stable state under this plasticity rule, it allows us to estimate the raw value of the contribution of rates and covariances in the evolution of synaptic weights. Here $B_{\alpha,\beta}$ can have any nonzero value, since it scales both the rate and covariance terms.
Under these conditions, our theory predicts that the rate term is at least an order of magnitude larger than the correlation term, but correlations can still have a small to moderate impact on the dynamics of synaptic weights (Fig. \ref{fig:DoCorrsMatterKohonen} \textbf{\textsf{D}}).

We next ask the opposite question: How do changes in synaptic weights impact firing rates, and covariances? The full theory  (see Eqs.~(\ref{eq:rbal}--\ref{eq:eqCorr}), and perturbative analysis in Materials and Methods) predict that the potentiation of \emph{EE} weights leads to large increases in rates and spike count covariances. This prediction was again confirmed by numerical simulations (Fig. \ref{fig:DoCorrsMatterKohonen} \textbf{\textsf{E}},\textbf{\textsf{F}}). 
This observation holds generally, and STDP rules that results in large changes in synaptic weights will produce large changes in rates and covariances.

Our theory thus shows that \emph{in general} weight dynamics can be moderately, but significantly affected by correlations when these are large enough (See Appendix~\ref{more_res_disc}\ref{hebbian_results} for a similar analysis on Classical Hebbian STDP). 
In turn, changes in synaptic weights will generally change the structure of correlated activity in a balanced network.


\begin{figure}[!htb]
\centering
  \includegraphics[width=\textwidth]{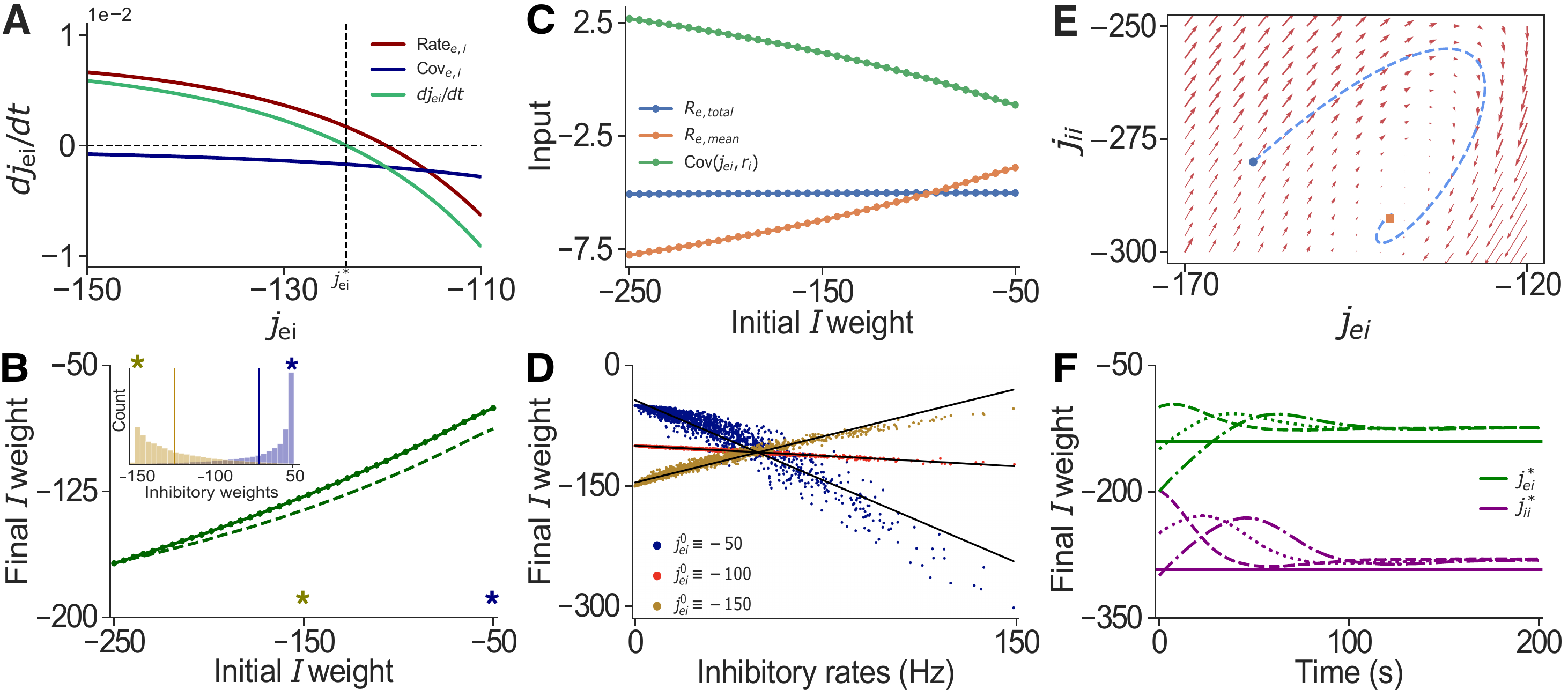}
\caption{\textbf{Correlations between weights and rates lead to a manifold of equilibrium weights.} \textbf{\textsf{A:}} The rate of change of the \emph{EI} weights as a function of the weights. The contributions of the covariance (blue) is considerably smaller than the contribution of the rate (red), and the system predicts a stable fixed point. \textbf{\textsf{B:}} A manifold of fixed points in $j_{\ee\ii}^*$--$j_{\ee\ii}^0$ space emerges due to correlations between weights and rates. Solid line represents simulations, dashed line are values obtained from the modified theory (Eqs.~(\ref{eq:eqCorr},\ref{eq:dJsoln},\ref{eq:rbal_mod})). Inset: Final distribution of \emph{EI} weights for a network with initial weights $j_{\ee\ii}^0=-150$ (yellow), and $j_{\ee\ii}^0=-50$ (blue). Different initial mean weights converge to different final mean weights.
\textbf{\textsf{C:}} Total recurrent input to \emph{E} neurons, $R_{\ee,\rm total} = \langle  \langle J_{kj}^{\ee\ee}r_j^e  \rangle_j+\langle J_{kl}^{\ee\ii}r_l^\ii \rangle_l \rangle_k$
for a range of initial weights. 
Mean recurrent input to \emph{E} cells is usually estimated by: $R_{\ee, \rm mean} = w_{\ee\ee} r_\ee + w_{\ee\ii} r_\ii$. The mean input deviates from the total input due to emergent correlations between weights and rates, ${\rm Cov}(J_{\ee\ii},r_\ii) = R_{\ee,\rm total} - R_{\ee, \rm mean} $.
\textbf{\textsf{D:}} The weights of individual \emph{EI} synapses corresponding to the same post--synaptic $E$ cell as a function of the equilibrium firing rates of pre--synaptic $I$ neurons. Each color represents a different simulation of the network with different initial \emph{EI} weight.
Equilibrium inhibitory weights and presynaptic rates are correlated (Blue: $R^2=0.952$, Red: $R^2=0.9865$, Yellow: $R^2=0.979$).
 \textbf{\textsf{E,F}} correspond to a system with inhibitory STDP in both \emph{EI} and \emph{II} synapses.  
 \textbf{\textsf{E:}} Phase portait of the $j_{\ee\ii}$--$j_{\ii\ii}$ system, with a sample trajectory (obtained solving Eqs.~(\ref{eq:rbal},\ref{eq:eqCorr},\ref{eq:JeiSoln}) iteratively over time) converging to the fixed point. 
 \textbf{\textsf{F:}} Sample trajectories of the $j_{\ee\ii}$--$j_{\ii\ii}$ system for a network of $N=10^4$ neurons in an asynchronous state. Simulations of three different initialization of inhibitory weights are shown using different dashed lines, while the theoretical prediction is shown with a solid line. Different initial weights converge to the same fixed point.}
\label{fig:eiVogels}
\end{figure}

\subsection{Balanced networks with inhibitory plasticity}

Next, we show that in its basic form our theory can fail in networks subject to inhibitory STDP,
and  how the theory can be extended to capture such dynamics.
The failure is due to correlations between weights and pre--synaptic rates which
are typically ignored~\cite{Vreeswijk1996, Vreeswijk1998, Rosenbaum2014,Landau2016,Pyle2016, Baker2019}, but can cause the mean--field description of network dynamics to become inaccurate.

To illustrate this, we consider a correlated balanced network subject to homeostatic plasiticity~\cite{Vogels2011}, which has 
been shown to stabilize the asynchronous balanced state and  conjectured to play a role in the maintenance of memories~\cite{Vogels2011,LitwinKumar2014,Hennequin2017}. 
Following~\cite{Vogels2011} we assume that \emph{EI} weights evolve according to

\begin{align} \label{eq:Jei}
\frac{dJ_{jk}^{\ee\ii}}{dt}= -\eta_{\ee\ii} \frac{J_{jk}^{\ee\ii}}{J^{\ee\ii}_0}\left[ (x_j^\ee-\alpha_\ee)S_k^\ii+ x_k^\ii S_j^\ee \right]
\end{align}

\noindent where $J^{\ee\ii}_0$ is the initial value of the $I$ to $E$ synaptic weight, which is uniform across synapses initially, and $\alpha_\ee$ is a constant that determines the target firing rates of $E$ cells. 
In a departure from the rule originally proposed by \emph{Vogels et al.}~\cite{Vogels2011}, we chose to  normalize the time derivative by its current and initial value, respectively.
This modification creates an unstable fixed point at zero, prevents \emph{EI} weights from changing signs, and keeps the analysis mathematically tractable (See Appendix~\ref{more_res_disc}\ref{modify_istdp} for details). 
The alternative hard bound at zero would create a discontinuity in the vector field of $J_{\ee\ii}$ complicating the analysis. 

Under the rule described by Eq.~\eqref{eq:Jei} a lone pre--synaptic spike depresses the synapse, while near--coincident pre-- and post--synaptic spikes potentiate the synapse. 
Changes in $EI$ weights steer the rates of individual excitatory cells to the target value $\rho_\ee:=\frac{\alpha_\ee}{2\tau_{STDP}}$. 
Indeed, individual \emph{EI} weights are potentiated if post--synaptic firing rates are higher than $\rho_\ee$, and depressed if the rate is below $\rho_\ee$. Our theory predicts that the network converges to a stable balanced state (Fig. \ref{fig:eiVogels}~\textbf{\textsf{A}}). Correlations again have only a mild impact on the evolution of synaptic weights (Fig. \ref{fig:eiVogels}~\textbf{\textsf{A}}).

Although our theory predicts a single stable fixed point for the average \emph{EI} weight, simulations show that weights converge to a different average depending on the initial \emph{EI} weights (Fig. \ref{fig:eiVogels}~\textbf{\textsf{B}} solid line). A manifold of stable fixed point emerges due to synaptic competition, which is due to heterogeneity in inhibitory firing rates in the network: The weights of highly active pre--synaptic inhibitory cells are potentiated more strongly  and more frequently, compared to those of lower firing cells (Fig. \ref{fig:eiVogels}~\textbf{\textsf{D}}). Thus while inhibitory rates and \emph{EI} weights are initially uncorrelated, correlations emerge as the excitatory rates approach their target.  
Network with different initial \emph{EI} synaptic weights, converge to different final distributions, and the emergent correlations between weights and rates drive the system to different fixed points (Fig. \ref{fig:eiVogels}~\textbf{\textsf{B}}). 

We used a semi--analytical approach to confirm that correlations between weights and rates explain the discrepancy between predictions of 
the mean field theory, and simulations. We can introduce a correlation dependent correction term into the  expression for the rates: 

\begin{align} \label{eq:rbal_mod}
\lim_{N\to\infty} \vec{r}=-\overline W^{-1}\big(\overline W_\xx r_\xx + \textrm{cov}(J_{\ee\ii},r_\ii) \big),
\end{align}

\noindent were $\textrm{cov}(J_{\ee\ii},r_{\ii}) := [ \langle \langle J_{jk}^{\ee\ii} r_{k}^{\ii} \rangle_{k}-\langle J_{jk}^{\ee\ii}\rangle_{k} \langle r^{\ii}_k\rangle_{k} \rangle_{j}, 0]^T$.
The  average covariances between weights and rates obtained numerically explain the departure from the mean--field predictions (Fig. \ref{fig:eiVogels}~\textbf{\textsf{C}}).  Using the corrected equation predicts mean equilibrium weights that agree well with simulations (Fig. \ref{fig:eiVogels} \textbf{\textsf{B}} dashed line).  

We next asked whether the accuracy of the mean--field description can be reestablished if the correlations between weights and rates are removed.  Such correlations are absent in a network with homogeneous inhibitory firing rates. Finding an initial distribution of weights that result in a balanced state with uniform inhibitory firing rates is non--trivial, and may not be possible outside of unstable regimes exhibiting rate--chaos where mean--field theory ceases to be valid~\cite{Ostojic2014}.
However, allowing \emph{II} synapses to evolve under the same plasticity rule we used for \emph{EI} synpases can homogenize inhibitory firing rates:
If we let

\begin{align} \label{eq:Jii}
\frac{dJ_{jk}^{\ii\ii}}{dt}= -\eta_{\ii\ii} \frac{J_{jk}^{\ii\ii}}{J_{\ii\ii}^0}\left[ (x_j^\ii-\alpha_\ii)S_k^\ii+ x_k^\ii S_j^\ii \right],
\end{align}

\noindent all inhibitory responses approach a target rate $\rho_\ii=\frac{\alpha_\ii}{2\tau_{STDP}}$, effectively removing the variability in $I$ rates. 
We conjectured that if inhibitory rates converge to a common target, synaptic competition would be eliminated,  and no correlations between weights and rates would emerge. This in turn would remove the main obstacle to the validity of a mean--field description. 

The evolution of the mean synaptic strengths under both \emph{II} and \emph{EI} plasticity is now given by Eq.~(\ref{eq:dJsoln}),

\begin{equation}  \label{eq:JeiSoln}
\begin{split}
\frac{ dJ_{\ee\ii}}{dt} = & -\eta_{\ee\ii} \frac{J_{\ee\ii}}{J^{\ee\ii}_0} \Big( (2\tau_{STDP} r_\ee -\alpha_\ee)r_\ii + 2\int_{-\infty}^\infty \widetilde K(f) \textrm{Re}[\langle S_\ee, S_\ii\rangle]df  \Big), \\
\frac{ dJ_{\ii\ii}}{dt}= & -\eta_{\ii\ii} \frac{J_{\ii\ii}}{J_{\ii\ii}^0} \Big( (2\tau_{STDP} r_\ii -\alpha_\ii)r_\ii + 2\int_{-\infty}^\infty \widetilde K(f) \langle S_\ii, S_\ii\rangle df  \Big).
\end{split}
\end{equation}

\noindent The fixed point of these equations can again be obtained using Eqs.~(\ref{eq:rbal},\ref{eq:eqCorr}) which predict that the network remains in a stable balanced state (asynchronous or correlated).
Note that $\eta_{\ii\ii}$ need not be greater than or equal to $\eta_{\ee\ii}$, since if $\eta_{\ee\ii} \gg \eta_{\ii\ii}$, then \emph{EI} weights will track the slow changes in \emph{II} weights, and the system will eventually converge to the stable fixed point.
These predictions agree with the results of simulations (Fig. \ref{fig:eiVogels}~\textbf{\textsf{E}},\textbf{\textsf{F}}). The stable manifold in the weights is replaced by a single stable fixed point, and the average weights and rates approach a state that is independent of the initial weight distribution. 

We next show that the balanced network subject only to \emph{EI} plasticity is robust to perturbatory inputs. Our theory predicts and simulations confirm that this learning rule maintains balance when non--plastic networks do not, and it can return the network to its original state after stimulation.


\begin{figure}[!htb]
\centering
    \includegraphics[width=\textwidth]{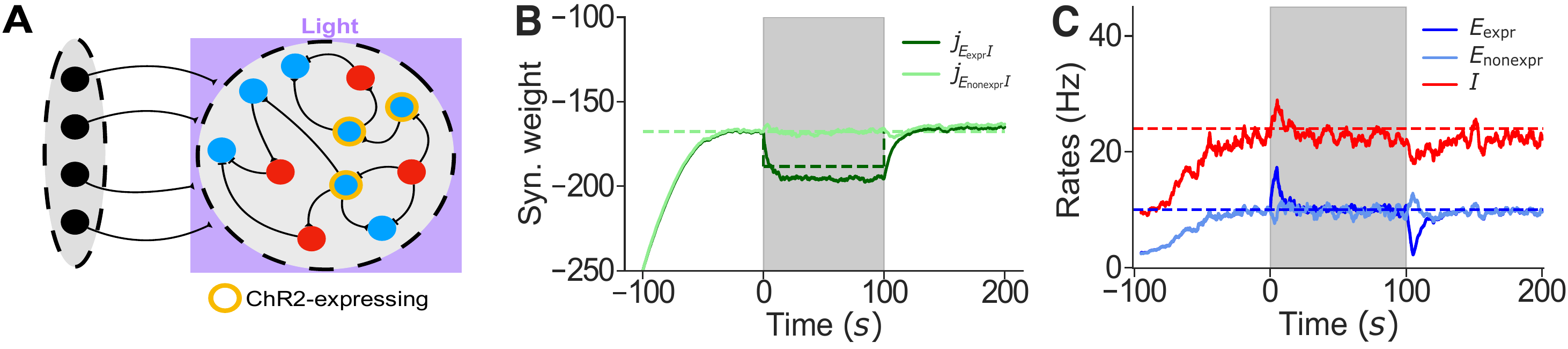}
\caption{\textbf{Homeostatic inhibitory plasticity maintains balanced in a stimulated network.} \textbf{\textsf{A:}} A recurrent network of excitatory, $E$, and inhibitory, $I$, neurons is driven by an external feedforward layer $X_1$ of correlated Poisson neurons. Neurons that express ChR2 are driven by optogenetic input, which is modeled as an extra layer of Poisson neurons denoted by $X_2$.
\textbf{\textsf{B:}} Time evolution of mean synaptic weights over the course of the experiment.
\textbf{\textsf{C:}} Mean firing rates over time. Inhibitory STDP maintains \emph{E} rates near the target, $\frac{\alpha_\ee}{2\tau_{\rm STDP}} $.
Solid lines represent simulations, dashed lines are values obtained from theory (Eqs.~(\ref{eq:rbal},\ref{eq:eqCorr},\ref{eq:Je1iSoln},\ref{eq:Je2iSoln})).}
\label{fig:Estim}
\end{figure}

\subsection{Inhibitory plasticity adapts response to stimuli}

Thus far, we analyzed the dynamics of plastic networks in isolation. However, cortical networks are constantly driven by
sensory input, as well as feedback from other cortical and subcortical areas. We next ask whether and how balance is restored if a subset of pyramidal neurons are stimulated~\cite{Ebsch2018}.  

In experiments using optogenetics not all target neurons express the channelrhodopsin 2 (ChR2) protein \cite{Adesnik2010,Boyden2005,Petrenau2007,Pouille2009}. 
Thus stimulation separates the target, \emph{e.g.} pyramidal cell population into stimulated and unstimulated subpopulations. 
Although classical mean--field theory produced singular solutions,  \emph{Ebsch, et al.} showed that the theory can be extended, and that  a non--classical balanced state is realized: Balance at the level of population averages ($E$ and $I$) is maintained, while balance at the level of the three subpopulations is broken. Since local connectivity is not tuned to account for the extra stimulation (optogenetics), local synaptic input cannot cancel external input to the individual subpopulations. However, the input averaged over the stimulated and unstimulated excitatory population is
cancelled. 

We show that inhibitory STDP, as described by Eq.~(\ref{eq:Jei}), can restore balance in the inputs to the stimulated and unstimulated. Similarly, \emph{Vogels, et al.} showed numerically that such plasticity restores balance in memory networks ~\cite{Vogels2011}.  Here, we peresent an accompanying theory that describes the evolution of rates, covariances, and weights  before, during, and after stimulation, and confirm the prediction of the theory numerically.

We assume that a subpopulation of  pyramidal neurons in a correlated balanced network receives a transient excitatory input. This could be a longer transient input from another subnetwork, or 
an experimentally applied stimulus.
To model this drive, we assume that the network receives input from two populations Poisson neurons,  $X_1$ and $X_2$. The first population drives all neurons in the recurrent  network, and was denoted by $X$ above.  The second population, $X_2,$ provides an additional input to a subset of excitatory cells in the network, for instance ChR2 expressing pyramidal neurons ($E_{\rm expr}$ in Fig. \ref{fig:Estim} \textbf{\textsf{A}}). The resulting connectivity matrix between the stimulated ($\ee_1$), unstimulated ($\ee_2$) and inhibitory ($\ii$) subpopulations, and the feed--forward input weight matrix have the form:

$$
\overline W=\left[\begin{array}{ccc}\overline w_{\ee_1\ee_1} & \overline w_{\ee_1\ee_2} & \overline w_{\ee_1\ii}\\ 
\overline w_{\ee_2\ee_1} & \overline w_{\ee_2\ee_2} & \overline w_{\ee_2\ii }\\
\overline w_{\ii \ee_1} & \overline w_{\ii\ee_2} & \overline w_{\ii\ii}\end{array}\right], \textrm{  and\   }  \overline W_x=\left[\begin{array}{cc}\overline w_{\ee_1\xx_1} & \overline w_{\ee_1\xx_2} \\
\overline w_{\ee_2\xx_1} & 0 \\ 
\overline w_{\ii \xx_1} & 0 \end{array}\right],
$$

\noindent where $\overline w_{ab}=p_{ab}j_{ab}q_b\sim\mathcal O(1)$, as before.  

The mean--field equation relating firing rates to average weights and input (Eq. \eqref{eq:rbal}) holds, with the vector of rates $r=[r_{\ee_1}, r_{\ee_2}, r_{\ii}]^T$ , and input vector $r_\xx = \left[ r_{\xx_1} ,    r_{\xx_2} \right]^T$. 
Similarly, mean spike count covariances are now represented by a $3\times 3$ matrix that satisfies Eq.~\ref{eq:eqCorr}. 
The  mean $E_1I$ and $E_2I$ weights evolved according to

\begin{align} \label{eq:Je1iSoln}
\frac{ dJ_{\ee_1\ii}}{dt}= & -\eta_{\ee_1\ii} \frac{J_{\ee_1\ii}}{J_{\ee_1\ii}^0} \Big( (2\tau_{\rm STDP} r_{\ee_1} -\alpha_\ee)r_\ii + 2\int_{-\infty}^\infty \widetilde K(f) \textrm{Re}[\langle S_{\ee_1}, S_\ii\rangle]df  \Big) \\ \label{eq:Je2iSoln}
\frac{ dJ_{\ee_2\ii}}{dt}= & -\eta_{\ee_2\ii}\frac{J_{\ee_2\ii}}{J_{\ee_2\ii}^0} \Big( (2\tau_{\rm STDP} r_{\ee_2} -\alpha_\ee)r_\ii + 2\int_{-\infty}^\infty \widetilde K(f) \textrm{Re}[\langle S_{\ee_2}, S_\ii\rangle]df  \Big).
\end{align}

We simulated a network of $N=10^4$ neurons in the correlated state with $c_{\xx_1}=c_{\xx_2}=0.1$. A subpopulation of $4000$ \emph{E} cells receives transient, correlated input. Solving Eqs.~(\ref{eq:Je1iSoln},\ref{eq:Je2iSoln}) predicts that inhibitory plasticity will modulate \emph{EI} synaptic weights so that the firing rates of both the $E_{\rm expr}$ and the $E_{\rm non-expr}$ approach the target firing rate before, during, and after stimulation. Once the network reaches steady state the mean inputs to each subpopulation cancel. Thus changes in \emph{EI} weights restore balance at the level of individual subpopulations or ``detailed balance,'' consistent with previous studies~\cite{Vogels2011,Hennequin2017}.
Simulations confirm these predictions (Fig. \ref{fig:Estim} \textbf{\textsf{B,C}}). When the input is removed, the inhibitory weights onto cells in the $E_{\rm expr}$ subpopulation converge to their pre--stimulus values, returning $E_{\rm expr}$ rates to the target value, and reestablishing balance (Fig. \ref{fig:Estim} \textbf{\textsf{B,C}}). 


We have thus shown that our theory captures the dynamics of stimulated plastic balanced networks. Homeostatic inhibitory STDP increases the stability and robustness  of balanced networks to perturbations by balancing inputs at a level of individual cells, maintaining balance in regimes in which non--plastic networks are unstable.
We presented an example in which only one subpopulation is stimulated. However, the theory can be extended to any number of subpopulations in asynchronous or correlated balanced networks receiving a variety of transient stimulus.


\section{Discussion}

We have developed an analytical framework that predicts the impact of a general class of STDP rules on the weights and dynamics of balanced networks. The balanced state is generally maintained under synaptic weight changes, as long as the rates remain bounded. 
Additionally, we found that correlations in spiking activity can introduce a small shift in the steady state, and change how quickly the fixed point is reached. 

We considered networks in the \emph{tight balance regime}~\cite{Ahmadian2019}. In this regime, large excitatory and inhibitory inputs cancel on average~\cite{Haider2006}, resulting in a fluctuation--driven state exhibiting irregular spiking. This cancellation is achieved when synaptic weights are scaled by $1/\sqrt{N}$ and external input is strong~\cite{Tsodyks1995, Vreeswijk1996, Vreeswijk1998, Renart2010}.  Our main assumption was that synaptic weights change slowly compared to firing rates. As this assumption holds generally,  we believe that our approach can be extended to other dynamical regimes. For instance supralinear stabilized networks SSNs operate in a loosely balanced regime where the net input is comparable in size to the excitatory and inhibitory inputs, and firing rates depend nonlinearly on inputs. Balanced networks and SSNs can behave differently, as they operate in different regimes. However, as shown in~\cite{Ahmadian2013}, SSN's and balanced networks may be derived from the same model under appropriate choices of parameters. In other words, the tight balanced solution can be realized in an SSN, and SSN--like solutions can be attained in a balanced network. This suggests that an extension of our general theory of plasticity rules to SSN's should be possible.

We obtained a mean--field description of the balanced network by averaging over the entire inhibitory and excitatory subpopulation, and a single  external population providing correlated inputs. As shown in the last section, the theory can naturally be extended to describe networks consisting of multiple physiologically or functionally distinct subpopulations, as well as multiple input sources. 

The mean--field description cannot capture the effect of some second order STDP rules as synaptic competition can correlate synaptic weights and pre--synaptic rates. We have shown that this can lead to different initial weight distributions converging to different equilibria. This can be interpreted as the maintenance of a previous state of the network over time. 

Mean--field theory can be extended to account for correlations between weights and rates. We were not able to find analytical expressions 
for these correlations, and therefore adopted a semi--analytic approach, which allowed us to explain the emergence of a manifold of steady--states. 
Finding an expression for  correlations between weights and rates would therefore lead to an accurate analytic extension of mean--field theory.  We have also shown that assuming \emph{II} weights evolve under a homeostatic STDP rule decorrelates weights and rates, resulting in a system that conforms to the predictions of classical mean--field theory.  

Partial stimulation of a population of \emph{E} neurons has been shown to break balance due to the incapacity of the network in cancelling inputs when weights are static~\cite{Ebsch2018}. I \emph{Ebsch et al} showed how classical balanced network theory can be modified to account for effects of the added input.
Additionally, inhibitory STDP has been shown to restore and maintain detailed balance in neuronal networks with stored patterns of memories or with tuned connectivity~\cite{Vogels2011}. However, inhibitory STDP was not tested in networks receiving optogenetic input, and studies of such STDP generally lacked an accompanying theory of the weight dynamics and focused on empirical simulations.
Our theory captures the dynamics of the network before, during, and after stimulation. In particular, our theory predicts that the network can remain balanced when plastic changes due to inhibitory STDP allow the network to cancel the optogenetic input.

\emph{Ocker et al} also studied the dynamics of mean synaptic weights in recurrent networks and derived equations for mean synaptic weights in terms of rates and covariances~\cite{Ocker2015}. This approach relied on linear response theory to obtain expressions of spike train covariances. Such a description is valid when neurons are driven by strong white noise in addition to weaker synaptic input from the network. 
Our results are thus complementary, describing networks in a different dynamical regime.

Excitatory plasticity has been linked to the formation of memories, assemblies, and cortical maps. However, in theoretical models, positive feedback loops created under these rules usually lead to unstable states if not constrained or controlled by homeostatic mechanisms~\cite{LitwinKumar2014,Zenke2015}. We showed that a balanced state can be attained when excitatory synapses follow Kohonen's rule, without fine tuning parameters and without imposing constraints on synaptic weights. Although inhibitory plasticity leads to robust, self--organized balance even in combination with commonly unstable excitatory Hebbian STDP~\cite{LitwinKumar2014}, its timescale is seconds or minutes; whereas evidence shows that such inhibitory rules operate on the timescale of hours or days~\cite{Kullmann2012,Vogels2013,Froemke2007,Zenke2015,Sprekeler2017,Hennequin2017}. Therefore, this framework provides mathematical tractability and biological relevance to a model of excitatory STDP \cite{Kohonen1984}, without the need for unrealistically fast inhibitory STDP.

The theoretical framework we presented is flexible, and can describe more intricate dynamics in circuits containing multiple inhibitory subtypes, and multiple plasticity rules, as well as networks in different dynamical regimes. Moreover, the theory can  be extended to plasticity rules that depend on third order  interactions~\cite{Pfister2006,Gjorgjieva2011}, such as the BCM rule~\cite{BCM1982}. This may produce richer dynamics, and change the impact of correlations.  


\section{Conclusion}

We developed a general second order theory of spike--timing dependent plasticity for classical asynchronous, and correlated balanced networks\cite{Vreeswijk1996,Vreeswijk1998,Renart2010,Baker2019}.  Assuming that synaptic weights change slowly, we derived a set of equations describing the evolution of firing rates, correlations as well as synaptic weights in the network. We showed that, when the mean--field assumptions are satisfied, these equations accurately describe the network's state, stability, and dynamics.  However, some plasticity rules, such as inhibitory STDP, can introduce correlations between synaptic weights and rates.  Although these correlations violate the assumptions of mean--field theory, we showed how to account for, and explain their effects. Additional plasticity rules can decorrelate synaptic weights and rates, reestablishing the validity of classical mean--field theory. Lastly, we showed that inhibitory STDP  allows networks to  maintain balance, and preserves the network's structure and dynamics when subsets of neurons are transiently stimulated.  Our approach is flexible and 
can be extended to capture interactions between multiple populations  subject to different plasticity rules.

\clearpage

\newpage

\section{Acknowledgments}
This work was supported by grants NIH-1R01MH115557 (KJ), NSF DMS-1654268 (RR) and DBI-1707400 (AA, RR, and KJ).

\bibliographystyle{apsrev4-1}
\bibliography{Manuscript_Refs.bib}

\appendix

\section{Review of mean--field theory in balanced networks} \label{rev_meanfield}

We consider recurrent networks of $N$ integrate--and--fire model neurons, $N_\ee$ of which are excitatory and $N_\ii$ inhibitory (with 
$N_\ee=0.8N$ and $N_\ii=0.2N$).  This population receives feed--forward synaptic input from an external population of $N_\xx$ excitatory 
neurons whose spike trains are modeled as Poisson processes, each having rate $r_\xx$ (Unless otherwise stated, in all simulations: 
$r_\xx=10$Hz). 
We use subscripts $\ee$, $\ii$ and $\xx$ for the local excitatory, local inhibitory, and external populations, respectively. 
The spike train of neuron $j=1,\ldots,N_a$ in population  $a=\ee,\ii,\xx$ is represented as a sum of Dirac delta functions,

$$
S_j^a(t)=\sum_n \delta(t-t_n^{a,j})
$$

\noindent where $t_n^{a,j}$ is the $n^{\rm th}$ spike time of the neuron. 
The synaptic input current to neuron $j=1,\ldots,N_a$ in population $a=\ee,\ii$ 
is given by

\begin{equation*}\label{E:Ij}
I^a_j(t)=R^a_j(t)+X^a_j(t)
\end{equation*}

\noindent where
 
$$
R^a_j(t)=\sum_{b=\ee,\ii}\, \sum_{k=1}^{N_b} J^{ab}_{jk} \sum_n \alpha_b(t-t_n^{b,k})
$$

\noindent is the recurrent input to the neuron from the excitatory ($b=\ee$) and inhibitory ($b=\ii$) neurons in the local circuit and 

$$
X^a_j(t)=\sum_{k=1}^{N_x} J^{ax}_{jk} \sum_n \alpha_x(t-t_n^{x,k})
$$

\noindent is the feed--forward input from the external population. 
Here, $J^{ab}_{jk}$ is the synaptic weight from presynaptic neuron $k$ in population $b$ to postsynaptic neuron $j$ in population $a$, and $\alpha_b(t)$ is a postsynaptic current (PSC) waveform, and is defined by

$$
\alpha_b(t)=\tau_b^{-1}e^{-t/\tau_b}H(t)
$$

\noindent where $H(t)$ is the Heaviside step function, and $\tau_b$ is the synaptic timescale of neurons in population $b=\ee,\ii,\xx$. In all simulations we take $\tau_\xx = 10$, $\tau_\ee = 8$, and $\tau_\ii = 4$~\cite{Baker2019}.
Without loss of generality, we assume that $\int \alpha_b(t)=1$. 

The membrane potential of neuron $j=1,\ldots,N_a$ in population $a=\ee,\ii$ obeys exponential integrate--and--fire dynamics~\cite{FourcaudTrocme2003},

$$
C_m\frac{dV_j^a}{dt}=-g_L(V_j^a-E_L)+g_L\Delta_T e^{(V_j^a-V_T)/\Delta_T}+I_j^a(t).
$$

\noindent See Table \ref{table:parameters} in Appendix~\ref{more_res_disc}\ref{parameters_sims} for a description of the parameters of the voltage dynamics. In our simulations, this equation was integrated using Forward Euler's method with step size $dt=0.1$.
We consider a random, blockwise--Erd\H{o}s--Renyi connectivity structure with

$$
J_{jk}^{ab}=\frac{1}{\sqrt N}\begin{cases}j_{ab} & \textrm{with prob. }p_{ab}\\ 0 & \textrm{otherwise,}\end{cases}
$$

\noindent where connections are statistically independent and $j_{ab},p_{ab}\sim\mathcal O(1)$ for $b=\ee,\ii,\xx$ and  $a=\ee,\ii$. Note that, as mentioned in the Discussion, our results are robust to different initial distributions of synaptic weights, and we provide code to implement them.

\subsection{Mean field theory of firing rates in balanced networks}\label{rates_meanfield}

We denote by $r_a$ the mean firing rate of neurons in population $a=\ee,\ii,\xx$, and let the components of the $2\times 1$ vector, 

$$
\vvec r=\left[\begin{array}{c}r_\ee\\ r_\ii\end{array}\right],
$$

\noindent
describe the mean activity of both subpopulations.
The mean--field synaptic inputs to neurons in populations $a=\ee,\ii$ are denoted by
$\overline U_a= \langle U_j^a(t) \rangle_j$
for $a=\ee,\ii$ and $U=I,X,R$. In vector form,
 
$$
\overline{\vvec U}=\left[\begin{array}{c}\overline{U_\ee} \\ \overline{U_\ii}\end{array}\right].
$$

\noindent We also define the recurrent and feed--forward mean--field connectivity matrices,

$$
\overline W=\left[\begin{array}{cc}\overline w_{\ee\ee} & \overline w_{\ee\ii}\\ \overline w_{\ii \ee} & \overline w_{\ii\ii}\end{array}\right], \quad \textrm{ and } \quad   \overline W_x=\left[\begin{array}{c}\overline w_{\ee\xx} \\ \overline w_{\ii \xx} \end{array}\right],
$$

\noindent where $\overline w_{ab}=p_{ab}j_{ab}q_b\sim\mathcal O(1)$.  Here, we have defined
$q_b=N_b/N$
which are assumed to be $\mathcal O(1)$. 

We review the mean--field analysis of firing rates in the balanced state~\cite{Vreeswijk1998}. The mean external input to each population is given by

$$
\overline{\vvec X}=\sqrt N \overline W_\xx r_\xx,
$$ 

\noindent and the mean recurrent input by 

$$
\overline{\vvec R}=\sqrt N \overline W \vvec r.
$$

\noindent The mean total synaptic input is therefore given by 

$$
\overline{\vvec I}=\sqrt N\left[\overline W \vvec r+\overline W_\xx r_\xx\right].
$$

In the balanced state, we have $\overline{\vvec I},\vvec r\sim\mathcal O(1)$, which can only be obtained by a cancellation between external and recurrent synaptic inputs. This cancellation requires $\overline W \vvec r+\overline W_\xx r_\xx\sim\mathcal O(1/\sqrt N)$ so that

\begin{equation}\label{eq:rbalance}
\lim_{N\to\infty} \vvec r=-\overline W^{-1}\overline W_\xx r_\xx
\end{equation}

\noindent in the balanced state, provided $\overline X_e /\overline X_i>w_{ei}/w_{ii}>w_{ee}/w_{ie}$ ~\cite{Vreeswijk1998}. The firing rates in Eq.~(\ref{eq:rbalance}) depend only on the recurrent structure and mean input, and are hence independent of the correlation structure in the network.

\subsection{Generating correlated spike trains} \label{corr_spikes}

In our simulations we generated spike trains, $S^\xx_j$, for the external population following \cite{Kuhn2003,Baker2019,Trousdale2012}. Here, we review the Multiple Interaction Process (MIP) algorithm for generating correlated Poisson processes. To generate $N_\xx$ processes, all having firing rate $r_\xx$ and pairwise correlation, $c_\xx$ we first generate a ``mother'' process with rate $r_m=r_\xx/c\xx$. Then, to generate each of the $N_\xx$ ``daughter'' processes, delete each spike from the mother process with probability $1-c_\xx$. In other words, each spike time from the mother process is included in each daughter process with probability $c_\xx$. Then the rates of the daughter processes are $r_m c = r_\xx$, as desired. Also,  two daughter processes share a proportion $c_\xx$ of spike times (since the probability that a spike time in one process also appears in the other process is $c_\xx$). 

This algorithm produces perfectly synchronous spikes. To make the spikes less synchronous, but still correlated, we ``jitter'' all of the daughter spike times by adding i.i.d. random variables to each daughter spike times~\cite{Trousdale2013}.

The cross spectral density (CSD) between two stationary processes is the Fourier transform of their cross--covariance function:

$$
\langle X,Y\rangle (f):=\int_{-\infty}^\infty c_{XY}(\tau)e^{-2\pi i f \tau}d\tau
$$

\noindent where

$$
c_{XY}(\tau):=\cov(X(t),Y(t+\tau)).
$$

\noindent We also define the mean--field spectra,

$$
\langle S_a,S_b\rangle=\textrm{avg}_{j,k} \langle S^a_k,S^b_k\rangle
$$

\noindent to be the average CSD between neurons in populations $a,b=\ee,\ii,\xx$ in the network. 

If the random variable (which is added to jitter spike times) has density $G(t)$ then the CSD is given by 

$$
\langle S_j^\xx,S_k^\xx\rangle(f)=\begin{cases}c r_\xx |\widetilde G(f)|^2 & j\ne k\\ r_\xx & j=k\end{cases}
$$

\noindent where $\widetilde G(f)$ is the Fourier transform of $G(t)$. 

In our simulations, we use Gaussian--distributed jitters with standard deviation $\tau_{\rm jitter}=5$ms, and we get 

$$
\langle S_j^\xx,S_k^\xx\rangle =c r_\xx e^{-4f^2\pi^2\tau_{\rm jitter}^2}
$$

\noindent if $j\neq k$.

\subsection{Mean--field theory of covariances in non--plastic balanced networks}\label{cov_meanfield}

Here we just state results already derived in \cite{Baker2019}. We define the mean--field spike count covariance matrix as,

$$
C=\left[\begin{array}{cc}C_{\ee\ee} & C_{\ee\ii}\\ C_{\ii \ee} & C_{\ii\ii}\end{array}\right]
$$

\noindent where $C_{ab}$ is the mean spike count covariance between neurons in populations $a=\ee,\ii$ and $b=\ee,\ii$ respectively.
In the balanced state for large enough $N$, this is given by 

\begin{equation}\label{eq:Covs}
C\approx \frac{1}{N} T_{\rm win} \overline W^{-1} \Gamma \overline W^{-T}
\end{equation}

\noindent where $T_{\rm win}$ is the size of the counting window used.  The matrix $\Gamma$ has the same structure as $C$ and represents the covariance between external inputs. 

If spike trains in the external layer are uncorrelated Poisson processes ($c_\xx=0$) then 

$$
\Gamma=\overline W_\xx \overline W_\xx^T\frac{r_\xx}{q_\xx}. 
$$

\noindent so that $C\sim\mathcal O(1/N)$, and the network operates in the asynchronous state \cite{Renart2010,Vreeswijk1998}.
If spike trains in the external network are correlated Poisson processes with pairwise correlation coefficient $c_\xx \neq 0$, then

$$
\Gamma=N\overline W_\xx \overline W_\xx^T c_\xx r_\xx
$$

\noindent so that $C\sim\mathcal O(1)$, and the network operates in a correlated state \cite{Baker2019}. 

These equations describe the expected spike count covariances within and between populations in the network.  

To simplify notation, we define the $2\times 2$ matrix

$$
\langle \vvec S,\vvec S\rangle=\left[\begin{array}{cc} \langle S_\ee,S_\ee\rangle & \langle S_\ee,S_\ii\rangle\\ \langle S_\ii,S_\ee\rangle & \langle S_\ii,S_\ii\rangle\end{array}\right].
$$

\noindent The generalization of the equations for spike count covariances to CSDs is

$$
\langle \vvec S,\vvec S\rangle=W^{-1} W_\xx \langle S_\xx,S_\xx\rangle W_\xx^* W^{-*}
$$

\noindent where $W_\xx^*$ is the conjugate transpose, and $W^{-*}$ is the inverse of the conjugate transpose,

$$
 W=\left[\begin{array}{cc} w_{\ee\ee} & w_{\ee\ii}\\ w_{\ii \ee} & w_{\ii\ii}\end{array}\right]\textrm{ and }   W_x=\left[\begin{array}{c}w_{\ee\xx} \\ w_{\ii \xx} \end{array}\right],
$$

$$
w_{ab}(f)=p_{ab}j_{ab}q_b\widetilde \alpha_b(f)\sim\mathcal O(1)
$$

\noindent and 

$$
\widetilde \alpha_b(f)=\frac{1}{1+2\pi i f \tau_b}
$$ 

\noindent is the Fourier transform of $\alpha_b(t)$. Note that $W(f)$ and $W_\xx(f)$ depend on frequency. 
Recall from the previous section that spike trains in the external population are correlated Poisson processes generated with Gaussian--distributed jittering, hence

$$
\langle S_\xx,S_\xx\rangle =c r_\xx e^{-4f^2\pi^2\tau_{c}^2}.
$$

To understand the relationship between these results and the spike count covariance results above, note that for stationary processes, the covariance between integrals over large time windows is related to the zero--frequency CSD according to

$$
\lim_{T\to\infty}\frac{1}{T}\cov\left(\int_0^T X(t)dt,\int_0^T Y(t)dt\right)=\langle X,Y\rangle(0).
$$

\noindent so that for large $T$, 

$$
\cov\left(\int_0^T X(t)dt,\int_0^T Y(t)dt\right)\approx T\langle X,Y\rangle(0).
$$

\noindent Now note that a spike count over a window of size $T$ is just an integral of spike trains over $[0,T]$ where $T=T_{\rm win}$. 
We can write the zero--lag covariance between two stationary processes in terms of the CSD as 

\begin{equation}\label{E:covXY}
\cov(X(t),Y(t))
=\int_{-\infty}^\infty \langle X,Y\rangle(f)df.
\end{equation}

\noindent This will be useful below when deriving an expression for the evolution of mean synaptic weights. 
More generally, the cross--covariance function, $c(\tau)=\cov(X(t),Y(t+\tau))$, is the inverse Fourier transform of $\langle X,Y\rangle(f)$. 

Some other useful properties of the cross--spectral operator are:

$$
\langle aX+bZ,Y\rangle=a\langle X,Y\rangle+b\langle Z,Y\rangle
$$

\noindent for $a,b\in\mathbb C$

$$
\langle X,Y\rangle =\langle Y,X\rangle^*
$$

\noindent where $z^*$ is the complex conjugate of $z$,

$$
\langle K*X,Y\rangle = \widetilde K\langle X,Y\rangle
$$

\noindent where $*$ denotes convolution, $K(t)$ is an $L^2$ kernel, and $\widetilde K(f)$ is the Fourier transform. 
The power spectral density of $X$ is given by
$
\langle X,X\rangle\in\R.
$

It is often useful to define the cross--spectral matrix between two multivariate processes. Specifically, suppose $\vec X(t)\in \R^m$ and $\vec Y(t)\in R^n$ are multivariate, stationary processes then $\langle \vec X,\vec Y\rangle\in \R^{m\times n}$ with

$$
[\langle \vec X,\vec Y\rangle]_{jk}=\langle \vec X_j,\vec Y_k\rangle.
$$

\noindent This operator has essentially all the same properties elucidated above except that the complex conjugate turns into a conjugate--transpose. 



\section{Accounting for the effects of correlations in the general STDP rule}\label{effect_corrs}

Recall that the eligibility trace, $x_j^a(t)$, of neuron $j$ in population $a$ evolves according to

\begin{equation} \label{eq:trace}
\tau_{\mathrm{STDP}}\frac{dx_j^a (t) }{dt}=-x_j^a (t) + S_j^a(t),
\end{equation}

\noindent for $a=\ee,\ii$, where  $S_j^a(t)=\sum_n \delta(t-t_n^{a,j})$ is the sequence of spikes of neuron, $j$.  The time constant, $\tau_{\mathrm{STDP}}$, defines the time over which concurrent spikes in two cells can lead to a change in synaptic weights. 

We assume that all synapses can be subject to activity--dependent modulations. The synaptic weight from neuron $k=1,\ldots,N_b$ in population $b=\ee,\ii$ to neuron $j=1,\ldots,N_a$ in population $a=\ee,\ii$ changes according to a generalized spike--timing dependent plasticity (STDP) rule,

\begin{align} \label{eq:dJ-supp}
\frac{dJ^{ab}_{jk}}{dt}=  \eta_{ab} \bigg(
A_0 +
\sum_{ \alpha=\{a,j\},\{b,k\}} A_{\alpha} S_\alpha +
\sum_{ \alpha,\beta=\{a,j\},\{b,k\}} B_{\alpha,\beta} x_\alpha S_\beta  \bigg)
\end{align}

\noindent where $\eta_{ab}$ is the learning rate that defines the timescale of synaptic weight changes, $A_0,A_\alpha,B_{\alpha\beta}$ are functions of the synaptic weight $J^{ab}_{jk}$
(since the change in the synaptic weight can depend on the current value of the synaptic weight in a linear or nonlinear manner), and $a,b=\ee,\ii$.  
For instance, the term $B_{(e,k),(i,j)}  x_k^\ee S_j^\ii$ describes the contribution due to a spike in post--synaptic cell $j$ in the inhibitory subpopulation, at the 
given value of the eligibility trace in the pre--synaptic cell $k$ in the excitatory subpopulation.

We assumed that changes in synaptic weights occur on longer timescales than the dynamics of the eligibility trace and the correlation timescale, \emph{i.e.} $1/\eta_{ab} \gg T_{\rm win},\tau_{\mathrm{STDP}}$. 
Let  $\Delta T$  be a time larger than the timescale of eligibility traces, $\tau_{\mathrm{STDP}},$ and $T_{\rm win}$, but smaller than $1/\eta_{ab}$, so that the time differential of the weights satisfies~\cite{Kempter1999}:

\begin{align}
\label{E:averaging-supp}
\frac{\Delta J^{ab}_{jk}}{\Delta T} & = \frac{\eta_{ab}}{\Delta T} \int_0^{\Delta T} \Big[  A_0 + \sum_{ \alpha=\{a,j\},\{b,k\}} A_{\alpha} S_\alpha +
\sum_{ \alpha,\beta=\{a,j\},\{b,k\}} B_{\alpha,\beta} x_\alpha S_\beta    \Big] dt.
\end{align}

We expand the sums and take split the integral over the terms in the sums,

\begin{align*}
\frac{\Delta J^{ab}_{jk}}{\Delta T} & = \frac{\eta_{ab}}{\Delta T} \bigg( \Delta T A_0 +
 A_{a,j} \int_0^{\Delta T} S^a_j  dt+ A_{b,k} \int_0^{\Delta T} S^b_k dt  \\
& + B_{\{ a,j\},\{ a,j\} } \int_0^{\Delta T}x^a_jS^a_j dt  + B_{\{ a,j\},\{ b,k\} } \int_0^{\Delta T} x^a_jS^b_k dt \\
& + B_{\{ b,k\},\{ a,j\} } \int_0^{\Delta T}x^b_kS^a_j dt+ B_{\{ b,k\},\{ b,k\} } \int_0^{\Delta T}x^b_kS^b_k dt  \bigg).
\end{align*}

\noindent Consider each integral that is multiplied by $1/\Delta T$ as a sample of joint statistics. Then:

\begin{equation}
\label{E:Jav-supp}
\begin{aligned}
\frac{\Delta J^{ab}_{jk}}{\Delta T} & =\eta_{ab} \bigg( A_0 +  A_{a,j} \mathrm{E}[S^a_j] + A_{b,k}) \mathrm{E}[S^b_k] +  B_{\{ a,j\},\{ a,j\} } \mathrm{E}[x^a_jS^a_j]  \\ 
& + B_{\{ a,j\},\{ b,k\} } \mathrm{E}[x^a_jS^b_k] + B_{\{ b,k\},\{ a,j\} } \mathrm{E}[x^b_kS^a_j]+ B_{\{ b,k\},\{ b,k\} }  \mathrm{E}[x^b_kS^b_k] \bigg)
\end{aligned}
\end{equation}

\noindent where $\mathrm{E}[\cdot]$ denotes expectation over time. 

We provide a detailed derivation for the term $B_{\{ a,j\},\{ b,k\} } \mathrm{E}[x^a_jS^b_k]$. The other terms are derived 
in the same way. First note that

$$
\mathrm{E}[x^a_jS^b_k]=\mathrm{E}[(K*S_j^a)(t)(S_k^b(t)]=\cov(K*S_j^a,S_k^b)+\mathrm{E}[K*S_j^a]\mathrm{E}[S_k^b]
$$

\noindent where $K(t)=e^{-t/\tau_{\rm{STDP}}}H(t)$ and $H(t)$ is the Heaviside function. 
The second term on the right hand side can be written in terms of the rates,

$$
\mathrm{E}[K*S_j^a]\mathrm{E}[S_k^b]=\tau_{STDP} r_j^a r_k^b
$$

\noindent where $r_j^a$ is the rate of neuron $j$ in population $a$ and  we used the fact that
$\int_{-\infty}^\infty K(t)dt=\tau_{STDP}.$
From Eq.~\eqref{E:covXY}, 

$$
\cov(K*S_j^a,S_k^b)=\int_{-\infty}^\infty \langle K*S_j^a, S_k^b\rangle(f)df.
$$

\noindent This can be simplified by recalling that convolutions interact nicely with CSDs 

$$
\langle K*S_j^a, S_k^b\rangle=\widetilde K \langle S_j^a, S_k^b\rangle
$$

\noindent where $\widetilde K(f)$ is the Fourier transform of the exponential kernel $K(t)$. This gives

$$
\cov(K*S_j^a,S_k^b)=\int_{-\infty}^\infty \widetilde K(f) \langle S_j^a, S_k^b\rangle(f)df.
$$

\noindent Therefore,

$$
\mathrm{E}[x^a_jS^b_k]= \int_{-\infty}^\infty \widetilde K(f)\langle S_j^a, S_k^b\rangle(f)df + \tau_{STDP} r_j^a r_k^b.
$$

\noindent The first term in the right hand side depends on the spike count covariance between spike trains of populations $a$ and $b$. The second term depends on firing rates of populations $a,b$.
Following the procedure demonstrated here and applying it to each term in Eq.~(\ref{E:Jav-supp}), we arrived to the following equation describing the evolution of weights,

\begin{align*}
\frac{\Delta J^{ab}_{jk}}{\Delta T} & = \eta_{ab} \bigg(A_0 + A_{a,j} r^a_j + A_{b,k} r^b_k\\
&  + B_{\{ a,j\},\{ a,j\} } \bigg( \int_{-\infty}^\infty \widetilde K(f)\langle S_j^a, S_j^a\rangle(f)df + \tau_{STDP} r_j^a r_j^a \bigg) \\
& +B_{\{ a,j\},\{ b,k\} }  \bigg( \int_{-\infty}^\infty \widetilde K(f)\langle S_j^a, S_k^b\rangle(f)df + \tau_{STDP} r_j^a r_k^b \bigg) \\
& + B_{\{ b,k\},\{ a,j\} } \bigg( \int_{-\infty}^\infty \widetilde K(f)\langle S_k^b, S_j^a\rangle(f)df + \tau_{STDP} r_k^b r_j^a \bigg) \\
& + B_{\{ b,k\},\{ b,k\} }  \bigg( \int_{-\infty}^\infty \widetilde K(f)\langle S_k^b, S_k^b\rangle(f)df + \tau_{STDP} r_k^b r_k^b \bigg) \bigg).
\end{align*}

\noindent Averaging both sides of the equation over all neurons $j$ and $k$ in populations $a$ and $b$, respectively, we obtained:

\begin{align*}
\frac{\Delta J_{ab}}{\Delta T} & = \eta_{ab} \bigg( A_0 +  A_{a,j} r_a + A_{b,k} r_b\\
& + B_{\{ a,j\},\{ a,j\} } \bigg( \int_{-\infty}^\infty \widetilde K(f)\langle S_a, S_a\rangle(f)df + \tau_{STDP} r_a r_a \bigg) \\
& + B_{\{ a,j\},\{ b,k\} } \bigg( \int_{-\infty}^\infty \widetilde K(f)\langle S_a, S_b\rangle(f)df + \tau_{STDP} r_a r_b \bigg) \\
& + B_{\{ b,k\},\{ a,j\} } \bigg( \int_{-\infty}^\infty \widetilde K(f)\langle S_b, S_a\rangle(f)df + \tau_{STDP} r_b r_a \bigg) \\
& + B_{\{ b,k\},\{ b,k\} }  \bigg( \int_{-\infty}^\infty \widetilde K(f)\langle S_b, S_b\rangle(f)df + \tau_{STDP} r_b r_b \bigg) \bigg).
\end{align*}

Lastly, we rearranged terms and arrived at a compressed expression dependent on rates and spike count covariances:

\begin{align} \label{eq:dJsoln-supp} 
\frac{d J_{ab}}{dt}  = \eta_{ab} \bigg( A_0 +  \sum_{\alpha,\beta=\{a,b\}}  \textrm{Rate}_{\alpha,\beta} + \textrm{Cov}_{\alpha,\beta} \bigg) ,
\end{align}

\noindent where

\begin{align*} 
\textrm{Rate}_{\alpha,\beta} &= A_{\alpha} r_{\alpha}/2 + B_{\alpha,\beta} \tau_{STDP}r_\alpha r_\beta \\
\textrm{Cov}_{\alpha,\beta} &= B_{\alpha,\beta} \int_{-\infty}^\infty \widetilde K(f)\langle S_\alpha, S_\beta\rangle(f)df,
\end{align*}

\noindent where all the coefficients remained the same, but changed notation, e.g. $B_{\{ a,j\},\{ b,k\} } = B_{ a,b }$.

\subsection{Implementation of inhibitory plasticity in numerical simulations}\label{implement_stdp}

We describe how we implemented the plasticity rules using the example of homeostatic inhibitory--to--excitatory STDP as in~\cite{Vogels2011}, which is a special case of our general STDP rule. Note that for the implementation of this rule, we will use the convention that $J_{jk}^{\ee\ii}<0$ or inhibitory weights.
After each presynaptic spike, $t_{k,n}^\ii$, of inhibitory neuron $k$, we make the update:

$$
J_{jk}^{\ee\ii}(t+\Delta t)=J_{jk}^{\ee\ii}(t)- \eta_{\ee\ii} \frac{J_{jk}^{\ee\ii}(t)}{J^{\ee\ii}_0} (x_j^\ee (t)-\alpha_\ee)
$$

\noindent and after each postsynaptic spike, $t_{j,m}^\ee$, of excitatory neuron $j$, we make the update:

$$
J_{jk}^{\ee\ii}(t + \Delta t)=J_{jk}^{\ee\ii}(t)- \eta_{\ee\ii} \frac{J_{jk}^{\ee\ii}(t)}{J^{\ee\ii}_0} x_k^\ii (t)
$$

\noindent where $J^{\ee\ii}_0$ is the initial weight from $\ii$ to $\ee$ neurons. Other STDP rules were implemented similarly. 

The rule from \cite{Vogels2011} was slightly modified to prevent $I$ to $E$ weights from becoming positive. In particular, the right hand side of the updates was multiplied by $\frac{J_{jk}^{\ee\ii}(t)}{J^{\ee\ii}_0}$, in order to create an unstable zero fixed point and enforce that weights will remain negative. See Appendix~\ref{more_res_disc}\ref{modify_istdp} for an example where inhibitory weights change signs if the zero fixed point is not present. 

This modification guarantees that \emph{EI} weights will remain negative in continuous--time rate--dynamics. However, \emph{EI} weights could still change sign, since in network simulations, changes in synaptic weights occur in spike--based discrete--time updates.
We thus proceed to derive a condition under which this modification ensures that synaptic weights will not change signs.
Consider updates due to post--synaptic spikes, since pre--synaptic spikes always strengthen the connection. We would like for the weights to always be negative:

$$
J_{jk}^{ei}=J_{jk}^{ei}-\frac{J_{jk}^{ei}}{J^{0}_{\ee\ii}} \eta_{\ee\ii} (x_j^e-\alpha_\ee)<0.
$$

\noindent Dividing through by $J_{jk}^{ei}$ and rearranging,

$$
-\frac{1}{J^{0}_{\ee\ii}} \eta_{\ee\ii} x_j^e + \alpha_\ee \eta_{\ee\ii} \frac{1}{J^{0}_{\ee\ii}} >-1.
$$

\noindent Since $J^{0}_{\ee\ii}<0$, it suffices to have:

$$
\alpha_\ee \eta_{\ee\ii} \frac{1}{|J^{0}_{\ee\ii}|} < 1
$$

\noindent Therefore, as long as this condition, which depends on network and plasticity parameters, is satisfied, $I$ to $E$ weights will not turn positive.

\subsection{Remarks on the general STDP rule}\label{remarks_stdp}

We would like to know what the impact of rates and covariances is in the synaptic weights. We estimate the magnitude of the two terms in the section below ``General impact of correlations in synaptic weights". The remaining question is: given the magnitude of each term, what is the qualitative effect that increasing rates or covariances will have on synaptic weights? Here, we show that changes in rates or covariances has two distinct effects: they can shift the location of the fixed point, or they can modulate the speed of convergence to that equilibrium.

Consider a general STDP rule that involves synapses between neurons of different populations. In this case, both the rates and covariances determine the location of the fixed point of the synaptic weights. In particular, increasing firing rates or spike count covariances can shift the location of the fixed point of the mean synaptic weight.

For simplicity, assume the STDP only involves second order terms (these give rise to covariance terms in the mean--field equation of the weights),

\begin{align*}
\frac{dJ^{ab}_{jk}}{dt} & = \eta_{ab} \bigg( B_{\{ a,j\},\{ a,j\} } x^a_jS^a_j + B_{\{ a,j\},\{ b,k\} } x^a_jS^b_k 
+ B_{\{ b,k\},\{ a,j\} } x^b_kS^a_j + B_{\{ b,k\},\{ b,k\} } x^b_kS^b_k \bigg) .
\end{align*}

\noindent As shown before, the mean synaptic weight evolves according to:

\begin{equation}
\label{eq:meanJproof}
\begin{aligned}
\frac{d J_{ab}}{dt} & = \eta_{ab} \bigg( B_{\{ a,j\},\{ a,j\} } \bigg( \int_{-\infty}^\infty \widetilde K(f)\langle S_a, S_a\rangle(f)df + \tau_{STDP} r_a r_a \bigg) \\
& +B_{\{ a,j\},\{ b,k\} } \bigg( \int_{-\infty}^\infty \widetilde K(f)\langle S_a, S_b\rangle(f)df + \tau_{STDP} r_a r_b \bigg) \\
& + B_{\{ b,k\},\{ a,j\} }  \bigg( \int_{-\infty}^\infty \widetilde K(f)\langle S_b, S_a\rangle(f)df + \tau_{STDP} r_b r_a \bigg) \\
& +B_{\{ b,k\},\{ b,k\} } \bigg( \int_{-\infty}^\infty \widetilde K(f)\langle S_b, S_b\rangle(f)df + \tau_{STDP} r_b r_b \bigg) \bigg)
\end{aligned}
\end{equation}

\noindent By assumption, $a\neq b$, and re--grouping terms, we get:

\begin{align*} 
\frac{d J_{ab}}{dt} & = \eta_{ab} \bigg( B_{\{ a,j\},\{ a,j\} }  \tau_{STDP} r_a r_a + (B_{\{ a,j\},\{ b,k\} }+B_{\{ b,k\},\{ a,j\} } ) \tau_{STDP} r_a r_b \\
&+ B_{\{ b,k\},\{ b,k\} }  \tau_{STDP} r_b r_b + B_{\{ a,j\},\{ a,j\} }  \int_{-\infty}^\infty \widetilde K(f)\langle S_a, S_a\rangle(f)df  \\
& +(B_{\{ a,j\},\{ b,k\} }+B_{\{ b,k\},\{ a,j\} } ) \int_{-\infty}^\infty \widetilde K(f) \textrm{Re}[\langle S_a, S_b\rangle(f)]df  \\
& + B_{\{ b,k\},\{ b,k\} }  \int_{-\infty}^\infty \widetilde K(f)\langle S_b, S_b\rangle(f)df \bigg)
\end{align*}

Grouping the first three terms and the last three terms we obtained:

\begin{align*} 
 \frac{d J_{ab}}{dt}  = \eta_{ab} \big( \textrm{Rate}_{a,b} + \textrm{Cov}_{a,b} \big)
\end{align*}

Thus rates and spike count covariances come additively into $dJ_{ab}/dt$. This implies that both rates and covariances determine the location of the fixed point of $J_{ab}$. Thus, increases in any of the two terms can lead to a shift in the location of the fixed point. In our framework, rates and covariances can be controlled independently through the $r_\xx$ and $c_\xx$, respectively. Therefore, increasing $c_{\xx}$ can put the network in regimes where $\textrm{Rate}_{a,b}$ and $\textrm{Cov}_{a,b}$ are of the same magnitude and hence increasing covariances can moderately shift the location of the fixed point $J_{ab}$. (This is the case for Kohonen's rule -- see Results in the  main text).

Let us now consider a general STDP rule that changes the weights of synapses between neurons of the same population, and these changes are only due to order 2 interactions. In this case, the rates and covariances do not determine the fixed point of the synaptic weights. However, rates and covariances can modulate the speed of convergence of the synaptic weights to steady state.

By assumption, the STDP only involves second order terms. Then:

\begin{align*}
\frac{dJ^{ab}_{jk}}{dt} & = \eta_{ab} \bigg( B_{\{ a,j\},\{ a,j\} } x^a_jS^a_j + B_{\{ a,j\},\{ b,k\} } x^a_jS^b_k 
+ B_{\{ b,k\},\{ a,j\} } x^b_kS^a_j + B_{\{ b,k\},\{ b,k\} } x^b_kS^b_k   \bigg)
\end{align*}

\noindent As shown before, the mean synaptic weight evolves according to Eq.~(\ref{eq:meanJproof}).
Since, $a=b$, and re--grouping terms, we get:

\begin{align*} 
 \frac{d J_{aa}}{dt} & = \eta_{ab} \bigg(B_{\{ a,j\},\{ a,j\} }+B_{\{ a,j\},\{ b,k\} }+B_{\{ b,k\},\{ a,j\} } + B_{\{ b,k\},\{ b,k\} } \bigg) \cdotp \\
 & \ \ \ \ \bigg( \tau_{STDP} r_a^2  + \int_{-\infty}^\infty \widetilde K(f)\langle S_a, S_a\rangle(f)df \bigg) \\
\end{align*}

\noindent Note that

$$
 \tau_{STDP} r_a^2  + \int_{-\infty}^\infty \widetilde K(f)\langle S_a, S_a\rangle(f)df > 0
$$

\noindent therefore, the only fixed point of $J_{aa}$ is determined by the roots of

$$
B_{\{ a,j\},\{ a,j\} }+B_{\{ a,j\},\{ b,k\} }+B_{\{ b,k\},\{ a,j\} } + B_{\{ b,k\},\{ b,k\} } =0.
$$

\noindent Thus the location of the fixed point of $J_{ab}$ is independent of rates and spike count covariances. However, since they come in multiplicatively in the derivative of $J_{ab}$ (\emph{i.e.}, the derivative is proportional to the sum of rate and covariance contributions), they can modulate the speed of convergence of the synaptic weights to the fixed point. This can happen through changes in $r_\xx$ or $c_\xx$.

\section{Additional Results and Discussion}\label{more_res_disc}

Here we include some results that support certain claims made throughout the paper. In particular, we show here that our theory holds for other STDP rules (not just Kohonen's, as demonstrated in the Results) such as Classical $EE$ Hebbian STDP.  We also show that in Classical Hebbian STDP increasing spike count covariances and dramatically increases the speed of convergence to equilbrium. And finally, we show that, \emph{in general}, the magnitude of rates in the evolution of synaptic weights is at least an order of magnitude larger than that of the spike count covariances.


\begin{figure}[!htb]
\centering
  \includegraphics[width=\textwidth]{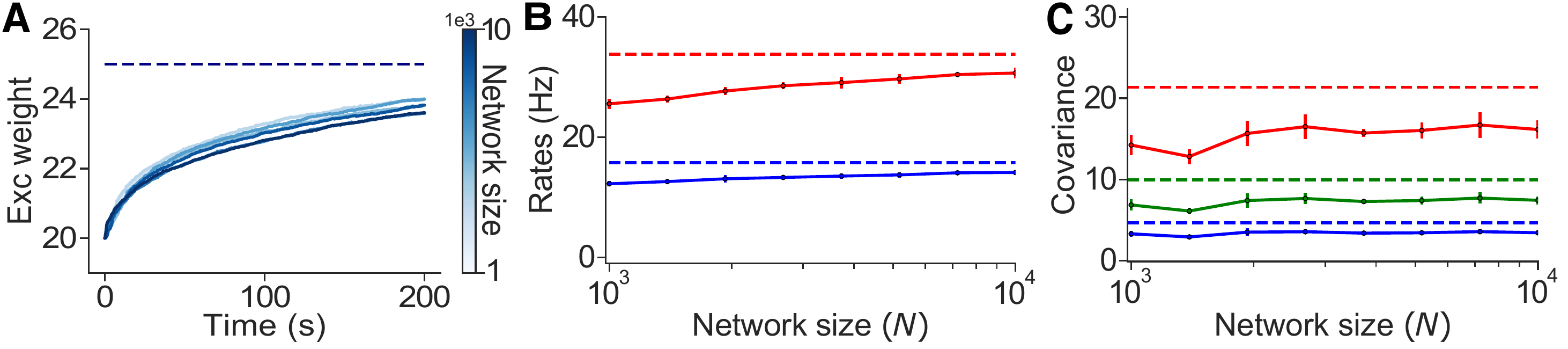} 
  \vspace{1mm}
\caption{\textbf{Classical Hebbian STDP yields a stable balanced state.} \textbf{\textsf{A:}} Mean excitatory synaptic weight, $j_{\ee\ee}$, evolving over time for increasing network sizes.
\textbf{\textsf{B:}} Mean $E$ and $I$ firing rates for increasing network sizes. \textbf{\textsf{C:}} Mean spike count covariance between $E$--$E$, $E$--$I$, and $I$--$I$ spike trains. Dashed lines represent the theory (Eqs.~(\ref{eq:rbalance},\ref{eq:Covs},\ref{eq:JeeSoln_supp})). Solid lines and dots are obtained from numerical simulations.}
\label{fig:AsymHebb}
\end{figure}

\subsection{Classical $EE$ Hebbian STDP}\label{hebbian_results}

\textbf{Asymptotic behavior.}
Assume that $EE$ synaptic weights evolve according to a Classical Hebbian STDP rule~\cite{Hebb1949,Markram1997,Bi1998} (See Table 1 in Materials and Methods), 

\begin{align} \label{eq:Jee_supp}
\frac{dJ_{jk}^{\ee\ee}}{dt} = \eta_{\ee\ee} \bigg( J_{\rm max}x_k^\ee S_j^\ee - J_{jk}^{\ee\ee}x_j^\ee S_k^\ee \bigg)
\end{align}

\noindent This rule yields the classical behavior original proposed by Hebb: a presynaptic spike followed by a postsynaptic spike potentiates the synapse. This equation also enforces that a postsynaptic spike followed by a presynaptic spike depresses the synapse. 

On average, the $EE$ synaptic weights evolve as:

\begin{align} \label{eq:JeeSoln_supp}
\frac{dJ_{\ee\ee}}{dt}= \eta_{\ee\ee} \bigg(J_{\rm max}-J_{\ee\ee} \bigg) \bigg(\tau_{\rm STDP}r_{\ee}^2 +\int_{-\infty}^\infty \widetilde K(f) \langle S_{\ee},S_{\ee} \rangle (f) df \bigg).
 \end{align}

The fixed point of Eq.~(\ref{eq:JeeSoln_supp}) is simply $J_{\ee\ee}^*=J_{\rm max}$ because Eqs.~(\ref{eq:rbalance},\ref{eq:Covs}) show that the second factor on the right hand side is strictly positive. In any case, we still solve for the firing rates and spike count covariances through their mean-field equations using the fixed point $J_{\ee\ee}^*=J_{\rm max}$.
 
Our theoretical framework predicts that the network attains a stable balanced state, and gives the location of the fixed point in terms of weights, rates, and covariances. We confirm our predictions with numerical simulations of the network in a correlated state (Fig. \ref{fig:AsymHebb}). 

Our theory (Eqs.~(\ref{eq:rbalance},\ref{eq:Covs},\ref{eq:JeeSoln_supp}) predicted that spike count covariances are order 1 and this is confirmed by numerical simulations (Fig. \ref{fig:AsymHebb} \textbf{\textsf{C}}).
As $N$ grows, firing rates and synaptic weights converge to the predicted values (Fig. \ref{fig:AsymHebb} \textbf{\textsf{A}},\textbf{\textsf{B}}).

We have now shown, that for $N$ large enough, the dynamics of a plastic balanced network can be well approximated by our theoretical framework for a range of plasticity rules derived from our general STDP rule, where the weights satisfy the balance condition.

\textbf{Impact of correlations in network dynamics.}
We have shown that in Kohonen's rule, increasing correlations mildly shifts the location of the fixed point. We now show a different effect of correlations in weight dynamics. Here, we use our theoretical framework to show that in a balanced network with $EE$ weights that change according to Eq.~(\ref{eq:Jee_supp}), increasing correlations does not affect the location of the fixed point, but increases the speed of convergence to that equilibrium.

\begin{figure}[!htb]
\centering
  \includegraphics[width=\textwidth]{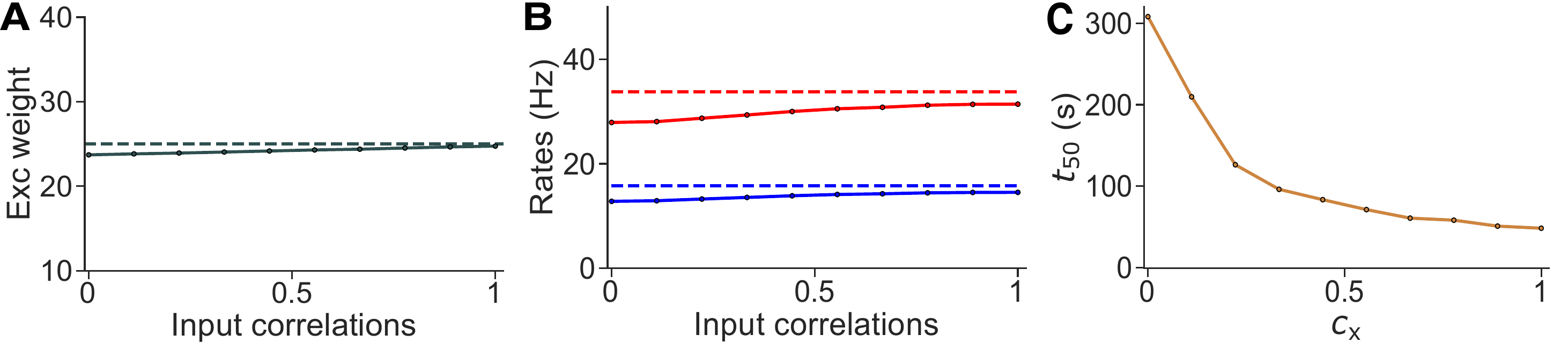}
    \vspace{1mm}
\caption{\textbf{Increasing input correlations modulates speed of convergence to equilibrium in Classical Hebbian STDP.} \textbf{\textsf{A:}} Mean excitatory synaptic weight, $j_{\ee\ee}$, obtained at different values of input correlations.
\textbf{\textsf{B:}} Mean $E$ and $I$ firing rates for increasing input correlations. \textbf{\textsf{C:}} Half time between initial condition and equilibrium for networks with increasing input correlations. Dashed lines represent the theory (Eqs.~\ref{eq:rbalance},\ref{eq:Covs},\ref{eq:JeeSoln_supp}). Solid lines and dots are obtained from numerical simulations.}
\label{fig:Corrs_Hebb}
\end{figure}

As mentioned before, the fixed point of synaptic weights does not change with increasing correlations (Eqs.~(\ref{eq:rbalance},\ref{eq:Covs},\ref{eq:JeeSoln_supp})) (Fig. \ref{fig:Corrs_Hebb} \textbf{\textsf{A}}). Since synaptic weights are not affected by this changes in correlations, rates also remain unchanged (Fig. \ref{fig:Corrs_Hebb} \textbf{\textsf{B}}). This is confirmed by numerical simulations. To measure how the speed of convergence is increased by larger correlations, we compute $t_{50}$ which is defined as the time it takes for the mean synaptic weight to reach 50\% convergence to the fixed point for increasing values of input correlations, and found that $t_{50}$ is dramatically reduced as correlations become larger, hence the mean weights converge faster and faster to equilibrium (Fig. \ref{fig:Corrs_Hebb} \textbf{\textsf{C}}).

We found that, as predicted in one of the previous sections (``Remarks on the general STDP"), the effect of correlations in the weight dynamics is different to that seen in Kohonen's rule. In this case, our theory predicts that increasing covariances (through our parameter $c_{\xx}$), speeds up the convergence to equilibrium, while maintaining the location of the fixed point unchanged.


\subsection{General impact of correlations in weight dynamics}\label{gen_corr_impact}

We have used our theory to explore the interaction between weights, rates, and covariances for a network with plastic synapses that followed Kohonen's rule and Classical $EE$ Hebbian STDP, and found that increasing correlations mildly impact the dynamics of the synaptic weights. We would now like to generalize the impact of correlations in the synaptic weights.

To do this, we estimate the covariance terms in our equation for the mean weights (Eq.~(\ref{eq:dJsoln-supp})) and compare them to the rate terms in that same equation. We assume there are interactions of order 2 only, and that these have constant coefficient. In other words, the synaptic weights evolve according to:

\begin{align*}
\frac{dJ^{ab}_{jk}}{dt} & = \eta_{ab} \big( x^a_jS^a_j + x^a_jS^b_k 
+ x^b_kS^a_j + x^b_kS^b_k \big)
\end{align*}

\noindent where we set all nonzero coefficients equal to one.
We then obtain an equation for the mean synaptic weights depending on the rates and covariances:

\begin{align}\label{eq:Jsoln_gen}
\frac{d J_{ab}}{dt}  = \eta_{ab}  \sum_{\alpha,\beta=\{a,b\}}  \textrm{Rate}_{\alpha,\beta} + \textrm{Cov}_{\alpha,\beta},
\end{align}

\noindent where
$
\textrm{Rate}_{\alpha,\beta} =  \tau_{STDP}r_\alpha r_\beta
$,
$\textrm{Cov}_{\alpha,\beta} =  \int_{-\infty}^\infty \widetilde K(f)\langle S_\alpha, S_\beta\rangle(f)df.
$
The fact that all nonzero coefficients are set to be equal allows us to estimate the raw values of the covariances and rates for each synapses (for $a,b=\ee,\ii$), and compare them.

\begin{figure}[!htb]
\centering
  \includegraphics[width=\textwidth]{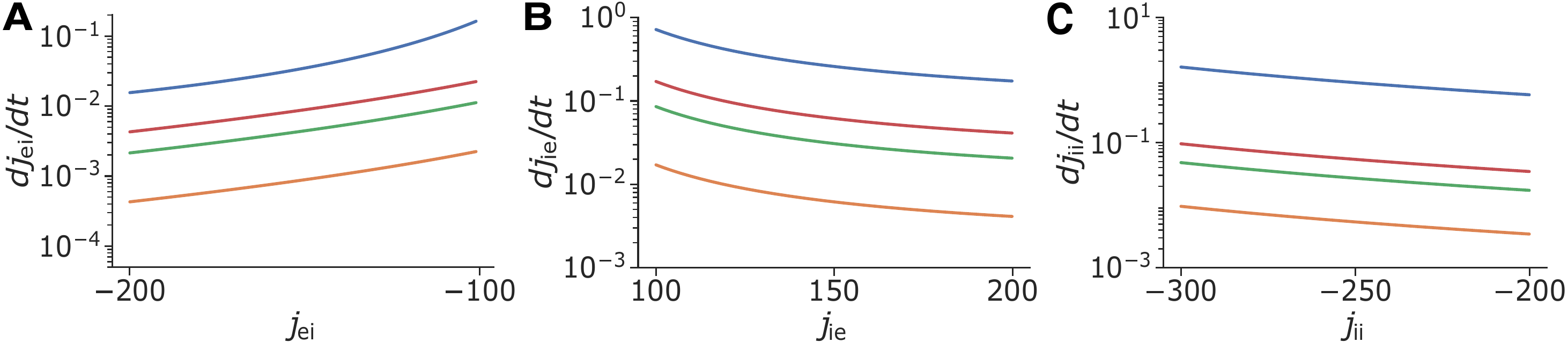}
  \vspace{1mm}
\caption{\textbf{General impact of correlations in synaptic weights.} \textbf{\textsf{A:}} Plot of the time derivative of the mean weight as a function of $j_{\ee\ee}$ assuming interactions of order 2 only.
\textbf{\textsf{B}} , \textbf{\textsf{C}}, \textbf{\textsf{D:}} Same as \textbf{\textsf{A:}} but for $EI$, $IE$, and $II$ syanpses, respectively. The contribution of rates to the evolution of synaptic weights is always larger than that of the covariances. Solid lines represent the theory  (Eqs.~\ref{eq:rbalance},\ref{eq:Covs},\ref{eq:Jsoln_gen}).}
\label{fig:generalCorr}
\end{figure}

We found that for a plasticity rule that satisfies our general rule (Eq.~(\ref{eq:dJsoln-supp})) acting on \emph{any} synapse, the impact of correlations in the synaptic weights is at least one order of magnitude smaller than the contribution of the rates to the dynamics of the weights (Fig. \ref{fig:generalCorr} \textbf{\textsf{A}},\textbf{\textsf{B}},\textbf{\textsf{C}},\textbf{\textsf{D}}). This implies that changes in the firing rates will usually have a stronger effect on the weights than changes in spike count covariances. However, increasing spike count covariances can still have a mild, but significant impact in synaptic weight dynamics when rates remain fixed. This last point is illustrated in our previous examples of networks undergoing excitatory plasticity (Kohonen and Classical Hebbian).


\begin{figure}[!htb]
\centering
  \includegraphics[width=\textwidth]{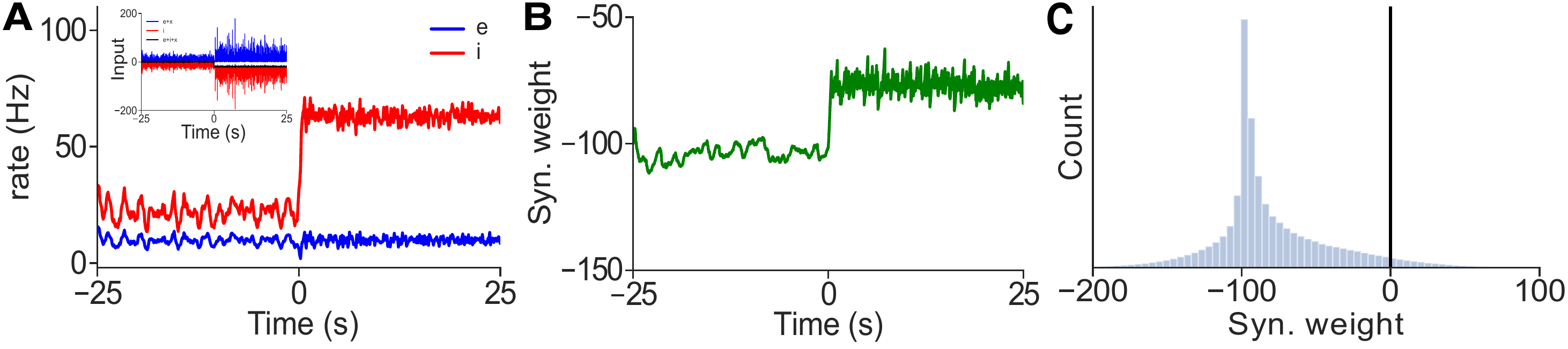}
    \vspace{1mm}
\caption{\textbf{Homeostatic nature of inhibitory STDP can break Dale's Law.} \textbf{\textsf{A:}} Excitatory and inhibitory firing rates over time. External stimulation starting at time 0sec increases the firing rate of \emph{I} cells. Inset: Average excitatory, inhibitory, and total inputs. Total input is negative when \emph{I} cells are stimulated.
\textbf{\textsf{B:}} Mean $EI$ synaptic weight over time. After stimulation onset, \emph{EI} weights decrease in magnitude. \textbf{\textsf{C:}} Distribution of $EI$ synaptic weights. A number of inhibitory synapses have turned positive. All plots obtained from a single simulation, where added external input to all inhibitory neurons starts at time 0s.}
\label{fig:Istim}
\end{figure}

\subsection{Modification to the inhibitory STDP rule}\label{modify_istdp}
We mentioned that the inhibitory STDP in~\cite{Vogels2011} was slightly modified to include a zero unstable fixed point in the synaptic weights. This was done because under certain relevant conditions, the synaptic weights can change sign in an effort to mantain $E$ rates at the target. 

One simple case in which this happens, is when all $I$ neurons are stimulated (Fig. \ref{fig:Istim}). The network is put in a semibalanced regime~\cite{Baker2020} where neurons receive excess inhibition on average (Fig. \ref{fig:Istim} \textbf{\textsf{A}} inset). The inhibitory rates increase, which would effectively decrease $E$ rates. However, $I$ to $E$ synaptic weights decrease such that the $E$ rates are mantained at the target (Fig. \ref{fig:Istim} \textbf{\textsf{A}},\textbf{\textsf{B}}). Since $EI$ weights are not controlled, excess inhibition causes some synapses to switch signs in order to mantain $E$ rates at the target (Fig. \ref{fig:Istim} \textbf{\textsf{C}}). This violates Dale's law, so in all other simulations of inhibitory STDP, the learning rule is modified to incorporate a zero unstable fixed point, which prevents synaptic weights to switch signs (as proven earlier in the Supporting Information).

\subsection{Technical details of simulations}\label{parameters_sims}
Neural networks described in the Materials \& Methods were simulated numerically using the set of parameters shown in Table \ref{table:parameters}. Code is available at \url{https://github.com/alanakil/PlasticBalancedNetsPackage}.

\begin{table}[h!]
  \begin{center}
 \begin{tabular}{|p{0.3\textwidth}|p{0.2\textwidth}|p{0.5\textwidth}|}
\hline
 \textbf{Connectivity parameter}
 & \textbf{Value}
 & \textbf{Description}
 \\
\hline
  $p_{ab}$ & 0.1 & Probability of connection for all $a,b=\ee,\ii$ 
 \\
\hline
   $j_{\ee\ee}/C_{m}$ & $25$mV & Weight of \emph{E} to \emph{E} synapses
 \\
\hline
   $j_{\ee\ii}/C_{m}$ & $-100$mV & Weight of \emph{E} to \emph{I} synapses
 \\
\hline
   $j_{\ii\ee}/C_{m}$ & $112.5$mV & Weight of \emph{I} to \emph{E} synapses
 \\
\hline
   $j_{\ii\ii}/C_{m}$ & $-250$mV & Weight of \emph{I} to \emph{I} synapses
 \\
\hline
\end{tabular}
\end{center}
\end{table}

 \begin{table}[h!]
  \begin{center}
 \begin{tabular}{|p{0.3\textwidth}|p{0.2\textwidth}|p{0.5\textwidth}|}
\hline
 \textbf{Neuron parameter}
 & \textbf{Value}
 & \textbf{Description}
 \\
\hline
  $C_{m}$ & 1 & Membrane capacitance 
 \\
\hline
   $g_{L}$ & $C_{m}/15$ & Leak conductance 
 \\
\hline
   $E_{L}$ & $-72$mV & Resting potential
 \\
\hline
   $V_{\rm th}$ & $-50$mV & Spiking threshold
 \\
\hline
   $V_{\rm re}$ & $-75$mV & Reset potential
 \\
\hline
   $\Delta_T$ & $1$mV & `Sharpness' parameter
 \\
\hline
   $V_T$ & $-55$mV & Threshold
 \\
\hline
\end{tabular}
\end{center}
\end{table}

\begin{table}[h!]
  \begin{center}
 \begin{tabular}{|p{0.3\textwidth}|p{0.2\textwidth}|p{0.5\textwidth}|}
\hline
 \textbf{Plasticity parameter}
 & \textbf{Value}
 & \textbf{Description}
 \\
\hline
  $\eta_{{ab}}$ & $10^{-4}$ & Learning rate of synaptic weights 
 \\
\hline
   $\tau_{\rm STDP}$ & $200$ms & Decay constant of eligibility traces
 \\
\hline
   $\rho_{\ee}$ & $10$Hz & Target rate of $E$ cells
 \\
\hline
   $\rho_{\ii}$ & $20$Hz & Target rate of $I$ cells
 \\
\hline
\end{tabular}
\end{center}
\caption{\textbf{Summary of simulation parameters.} 
} \label{table:parameters}
\end{table}

%
%
%



\end{document}